\documentclass[arxiv]{aastex631}

\usepackage{amsmath}

\shorttitle{Gaussian Process Blend Identification}
\shortauthors{Buchanan et al.}

\graphicspath{{./}{figures/}}

\begin{document}

\title{Gaussian Process Classification for Galaxy Blend Identification in LSST}

\correspondingauthor{James J. Buchanan}
\email{buchanan11@llnl.gov}

\author[0000-0001-8207-5556]{James J. Buchanan}
\affiliation{Physics Division, Lawrence Livermore National Laboratory, Livermore, CA 94550, USA}

\author[0000-0002-8505-7094]{Michael D. Schneider}
\affiliation{Physics Division, Lawrence Livermore National Laboratory, Livermore, CA 94550, USA}

\author{Robert E. Armstrong}
\affiliation{Physics Division, Lawrence Livermore National Laboratory, Livermore, CA 94550, USA}

\author[0000-0002-9787-1392]{Amanda L. Muyskens}
\affiliation{Computational Engineering Division, Lawrence Livermore National Laboratory, Livermore, CA 94550, USA}

\author[0000-0003-3806-7369]{Benjamin W. Priest}
\affiliation{Center for Applied Scientific Computing, Lawrence Livermore National Laboratory, Livermore, CA 94550, USA}

\author[0000-0001-6368-3791]{Ryan J. Dana}
\affiliation{Global Security Computing Applications Division, Lawrence Livermore National Laboratory, Livermore, CA 94550, USA}

\begin{abstract}
A significant fraction of observed galaxies in the Rubin Observatory Legacy 
Survey of Space and Time (LSST) will overlap at least
one other galaxy along the same line of sight, in a so-called ``blend.''
The current standard method of assessing blend likelihood in LSST images
relies on counting up the number of intensity peaks in the
smoothed image of a blend candidate, but the reliability of this procedure has
not yet been comprehensively studied. Here we construct a realistic
distribution of blended and unblended galaxies through high-fidelity 
simulations of LSST-like images, and from this we examine the blend 
classification accuracy of the standard peak-finding method. Furthermore, we develop
a novel Gaussian
process blend classifier model, and show that this classifier is competitive
with both the peak-finding method as well as with a convolutional neural
network model. Finally, whereas the peak-finding method does not naturally
assign probabilities to its classification estimates, the Gaussian process
model does, and we show that the Gaussian process classification probabilities
are generally reliable.
\end{abstract}


\section{Introduction} \label{sec:intro}
The Vera C. Rubin Observatory, under construction, will undertake the 10 year Legacy Survey of Space and Time (LSST; \citealt{2019ApJ.873.2}) beginning in 2023. It will pursue numerous scientific objectives including a wide-field study of weak lensing shear, the correlated but relatively minuscule distortion of multiple galaxy images attributable to gravitational lensing by a common intervening matter concentration. This informs estimates of the distribution of matter in the universe \citep{2015RPP.78.086901}, serving as a probe of, e.g., dark energy \citep{2018_DESC_SRD}. Weak lensing correlations are statistically inferred by associating the measured shapes of multiple galaxies, meaning that weak lensing estimates of the cosmological matter distribution generally improve with the number of well-measured galaxy shapes as long as systematic errors can be controlled \citep{2018ARAA.56}. The LSST will thus attempt to observe and characterize the properties of an unprecedented number of galaxies throughout a wide and deep field of view.

Due to the survey depth, a significant fraction of observed galaxies will overlap at least one other galaxy along the same line of sight, resulting in a so-called ``blend.'' Among objects detected in the Hyper Suprime-Cam (HSC) Wide survey, 58\% are part of the same contiguous region of highly luminous pixels as other objects \citep{2018PASJ.70.SP1.a}. The LSST's greater depth will exacerbate this issue still further: \citet{2021JCAP.2021.07} estimate that 62\% of galaxies detected in the full LSST survey depth will overlap with another source over at least 1\% of their total flux. Multiple galaxies should not be attributed to a single fictitious super-galaxy, so correctly identifying blends is important. Meanwhile, single galaxies should not be separated into multiple fictitious sub-galaxies, i.e.\ incorrectly designating a singleton as a blend is to be discouraged. In cases where this decision cannot be made with confidence \citep{Dawson_2015}, the corresponding image regions could be omitted from subsequent analysis. However, doing so reduces the statistical power of the survey and introduces difficult-to-quantify selection biases, so accurate decisions must be made as often as possible. This requires both a maximally discriminating blend identification procedure as well as a precise estimate of its uncertainty. In this paper, we develop a Gaussian process model for probabilistic identification of blends addressing both of these concerns.

The properties of each individual galaxy in a blend must then be inferred by an automated procedure called a ``deblender.'' The deblender in the HSC software pipeline \citep{2018PASJ.70.SP1.a} and the SCARLET framework \citep{2018AandC.24}, the default deblender in Rubin's LSST Data Management Science Pipelines software stack\footnote{\href{https://pipelines.lsst.io}{pipelines.lsst.io}} (LSST Science Pipelines), both require an initial estimate of the total number of galaxies participating in a given blend. In the current versions of both pipelines, this estimate is made by counting up the number of intensity peaks in a smoothed image of the neighborhood around a blend. However, while this procedure has been studied in the context of HSC images \citep{2018PASJ.70.SP1.a}, its reliability for LSST images has not yet been comprehensively examined. The two surveys differ in several respects relevant to the problem of galaxy separation. The greater depth of LSST compared to the HSC survey will exacerbate the blending problem as remarked above, and the dimmest galaxies detected by LSST will generally be more highly redshifted than typical galaxies in HSC. In addition, the general appearance of galaxy images will differ between the two survey instruments due to e.g.\ different sky coverage widths per pixel ($\sim 1/6$\,arcsec for HSC vs. $1/5$\,arcsec for LSST), different atmospheric characteristics at the telescope sites, and different instrumental throughputs between distinct filter bands. Thus, in the present study we begin to characterize the performance of peak finding for blend identification in the specific context of the LSST, by constructing a realistic set of simulated LSST images containing tens of thousands of visible galaxies with realistic distributions of location, orientation, brightness, and other properties, and then testing the reliability of peak finding on those images.

Meanwhile, the method of peak finding does not naturally assign probabilities to its estimates. It yields a single ``best-guess'' number of galaxies in a given image region. If a deblender like those listed above is given just that best guess, then it will produce just one possible deblending result, with no direct indicator of its likelihood. Because deblenders cannot always reconstruct the properties of highly blended galaxies with high fidelity, it may be preferable to ignore some likely blends rather than attempt to deblend them. This is most straightfowardly implemented by putting a cut on the probability of blending in a given image region, but peak finding does not readily admit such a strategy. In contrast, the Gaussian process model developed here comes equipped with a full set of estimates of blend probabilities. These estimates are comprehensively informed by the degree of similarity of observed data to training examples, so that in cases where some observed light profile is very dissimilar to anything seen before, the model will correspondingly report a high degree of uncertainty, and vice versa. Furthermore, after training the Gaussian process model on a set of simulated galaxy images, its blend identification accuracy on previously unseen images equals or exceeds that of the peak-finding method in almost every context we examine. On the other hand, peak finding naturally yields estimates of the centroids of each astronomical object, whereas the other blend identification models we study here do not. Furthermore, peak finding produces a count of all significantly distinguishable objects in a blend, whereas our blend identification classifier models only produce a single probability of ``blendedness.'' In a complete image processing pipeline, additional model expressiveness beyond a single blendedness value would be required in order to estimate a specific number of objects in each blend along with their locations. We propose one such model architecture here, as a basis for future study.

In this study we have provisionally assumed that object detection, counting and localization, and deblending will be performed as distinct steps in a fixed sequence, corresponding to the current implementation of the LSST Science Pipelines. In contrast, a number of proposed convolutional neural network (CNN) models relax this assumption, since a CNN naturally accepts an image pixel grid as input with minimal preprocessing. The CNN model of \citet{2021_wakesleep} performs detection, centering, and multiband flux measurement of stars in a crowded stellar field imaged by the Sloan Digital Sky Survey (SDSS; \citealt{2012APJS.203.2}). The result is a fully probabilistic catalog, in which both the properties and the overall number of objects in an image are assigned posterior probability distributions, requiring only a multiband astronomical image as input. However, that specific model fails to adequately handle galaxies, which, unlike stars, can have a variety of different sizes and shapes in a typical image. The Mask R-CNN network of \citet{2019MNRAS.490.3}, developed and tested using real and simulated Dark Energy Camera (DECam) images \citep{2019AJ.157.5}, performs detection, deblending, and star vs. galaxy classification simultaneously, given the original multiband image as input. That paper illustrates the need to tune models for specific survey depths and to re-examine those models when the depth changes: the performance of the Mask R-CNN model trained on a shallower image set drops significantly when applied to a deeper, more highly redshifted image set. The performance of any of the aforementioned CNN models on LSST-like images remains to be seen.

Given a simulated, multiband LSST image and an estimate of each galaxy's centroid as inputs, \citet{2020MNRAS.500.1} use CNNs in a variational autoencoder to deblend and denoise individual galaxies. As the authors note, the performance of that network depends on the quality of the galaxy centroid estimates. Peak finding is a standard algorithm for this task, but it does not always select a point very close to a true galaxy centroid in cases of unrecognized blends \citep{Dawson_2015}. The same concern arises for any other model that requires an initial estimate of object locations as input, including SCARLET, a variant of SCARLET developed by \citet{2019_pixelcnnscarlet} incorporating a PixelCNN-based galaxy morphology prior, and the branched GAN architecture of \citet{2019MNRAS.485.2}. Correctly recognizing a larger fraction of blend instances could thus help to flag some of the less reliable deblending results from these sorts of models. In the present study, we evaluate a simple CNN classifier trained for the specific task of blend identification on LSST \textit{i}-band images, comparing its performance to an even simpler logistic regression classifier, the peak-finding method, and the Gaussian process classifier. Most of the CNN studies listed above evaluate their model performance on simulated images of galaxies with somewhat arbitrary distributions of visible properties. In contrast, here we test our models on a population of simulated galaxies with realistically correlated distributions of apparent sizes, morphologies, luminosities, and positions in the sky, resulting in estimates of model performance with comparatively straightforward astronomical significance.

Gaussian process models have previously been developed for earlier and later steps in a typical image processing workflow. \citet{2021_GPimpute} use Gaussian process regression to impute missing pixel values in astronomical images. \citet{2021AJ.162.3} and \citet{2021AandA.650.A81} explore different Gaussian process regression models for interpolating astrometric calibrations in DECam and HSC images, respectively. The specific Gaussian process model used in the present study is very similar to that developed by \citet{2021_MuyGPs_StarGalaxy} to classify isolated stars vs. galaxies in HSC images. Also like the present study, the model of \citet{2021_MuyGPs_StarGalaxy} is tuned and applied using the MuyGPs software package \citep{2021_MuyGPs}. By way of comparison, here we use different hyperparameter settings tuned for the specific task of galaxy blend identification on simulated LSST images, and we also explore a distinct approach to uncertainty quantification by examining the probabilistic calibration of our model predictions.

The remainder of this paper is structured as follows: Section~\ref{sec:imagesim} describes the underlying galaxy distribution and the procedure used to simulate images of these galaxies as they would be seen in the LSST. Section~\ref{sec:preprocessing} details how a telescope image is transformed into model-ready input. Sections~\ref{sec:peak_finding} and \ref{sec:gp_model} describe the peak-finding and Gaussian process classifier models, respectively, and report their classification accuracy on the simulated image data. Section~\ref{sec:comparison} compares the strengths and weaknesses of these and other models. The fidelity of a probabilistic interpretation of these models is examined in section~\ref{sec:uq}. Finally, section~\ref{sec:conclusions} summarizes the key results and establishes an agenda for further research.

\section{Image Simulation}\label{sec:imagesim}
The basic inputs to the models considered here are a set of simulated images of a simulated galaxy distribution, using catalogs produced by the LSST Dark Energy Science Collaboration (LSST DESC; \citealt{2012_DESC}). Simulated galaxy properties are taken from the cosmoDC2 v1.1.4 catalog \citep{2019APJS.245.2} and the DESC DC2 Truth Match dr6 v1 catalog \citep{2021_DESCDC2_datarelease}. The distribution of galaxies in the cosmoDC2 catalog is based on the Outer Rim run \citep{2019APJS.245.1}, a trillion-particle, $(4.225\,\text{Gpc})^3$ simulation using a cosmological model based on Wilkinson Microwave Anisotropy Probe data \citep{2011APJS.192.2}. Galactic light cones were modeled out to a depth of $z = 3$ and the catalog contains a realistic number density of galaxies up to an \textit{r}-band magnitude of 28, along with many additional galaxies at higher magnitudes. This and numerous other cosmoDC2 statistics have been confirmed \citep{2019APJS.245.2} to be in close agreement with observed data from a variety of sources including the HSC \citep{2018PASJ.70.SP1.b}, COSMOS \citep{2007APJS.172.1}, DEEP2 \citep{2013APJS.208.1}, and SDSS \citep{2017APJS.233.2}.

CosmoDC2 is the observing target of the LSST DESC DC2 Simulated Sky Survey \citep{2021APJS.253.1}, a comprehensive simulation of LSST observations covering a roughly square-shaped, $300\,\text{deg}^2$ region of the southern sky, with a total number of exposures corresponding to half of the total survey duration. The DESC DC2 Truth Match catalog contains the true values of certain measurable properties of galaxies imaged in the survey, including their total fluxes integrated over each filter bandpass, accounting for the effects of internal reddening, redshift, and atmospheric and instrumental throughputs. For the image simulation performed here, these fluxes as well as the gravitationally lensed R.A.\ and decl.\ for each galaxy were taken from this catalog. Other quantities for each galaxy --- weak lensing convergence and shear components, the fraction of total flux concentrated in the galactic bulge, and the separate Sersic index, size, and ellipticity components for the galactic disk and bulge --- were taken from the cosmoDC2 catalog.

Images were simulated in GalSim \citep{2015AandC.10}. Images take a desired image location within the DC2 coverage area and tabulate all catalog galaxies in a window around that location large enough to fully encompass the desired image size, $2048^2$ pixels, with a resolution of 0.2\,arcsec per pixel. For each tabulated galaxy, separate Sersic profiles for the galactic disk and bulge are defined in GalSim and transformed to match the catalog values of lensed R.A.\ and decl., unlensed ellipticity, weak lensing shear, and convergence. The full DESC DC2 image simulation includes the addition of random ``knots'' of star formation to each galaxy, with knot characteristics correlated to galaxy properties in such a way as to further enhance the realism of the galaxy light profiles. We have not attempted to replicate that effort here. We expect that this has a subleading impact on the present analysis of Gaussian process and peak-finding model performance. A preliminary examination using the light profiles produced in the full DESC DC2 simulation indicates that the performance of the Gaussian process model is largely unaffected, which is also consistent with the relative insensitivity of this model to the specific footprint construction method (as shown in section~\ref{sec:gp_model}). This insensitivity is in large part due to the use of principal component analysis (PCA) dimensionality reduction on galaxy image cutouts (described in section~\ref{sec:preprocessing}). This has the effect of prioritizing large-scale features and ignoring smaller-scale variations, and marginal increases in image realism generally just affect the latter. Since smoothing the image (as described in section~\ref{sec:peak_finding}) strongly subdues light profile variations over angular scales smaller than typical galaxy sizes, the peak-finding method should be similarly insensitive to this issue. However, a CNN classifier may potentially suffer a relatively significant drop in accuracy on real images if training images are not highly realistic (see, e.g., \citet{2019MNRAS.490.4}). Future work will look to incorporate the light profiles from the full DESC DC2 simulation.

Based on the single-image exposure time (30\,s), the telescope mirror diameter (effectively 642.3\,cm), and expected LSST atmospheric and instrumental throughputs, the flux value from the Truth Match catalog is converted to a mean expected number of photons arriving at the camera CCDs, and the overall intensity of each galaxy is set to this value. The whole scene is randomly shifted by up to 0.2\,arcsec (one pixel width) horizontally as well as vertically, which simulates the net result of a random translational dither. The sum of the profiles of all galaxies in an image then defines a probability distribution for photon emission that is used as the basis for a simple photon-shooting simulation, starting from ``the top of the atmosphere.'' The photon trajectories are perturbed by an isotropic Kolmogorov point spread function (PSF) with a typical full width at half maximum of 0.7\,arcsec (this value was randomly varied by around 10\% in each exposure). The number of photons arriving in a given pixel results in an equal number of converted electrons released into the pixel volume, whose trajectories are modeled using GalSim's {\tt lsst\_itl\_32} silicon sensor simulation. A gain of 1 is assumed, so that every photoelectron converts into exactly 1 analog-to-digital unit read out by the sensors. On top of this, for every exposure a night sky background level is selected, with a mean value matching an estimated dark sky magnitude of 20.47 at the telescope site, randomly varying by around 5\% in each exposure. This is converted into a number of photons per pixel, which is used as the mean of a Poisson-distributed intensity added to each pixel.

\begin{figure}
\plotone{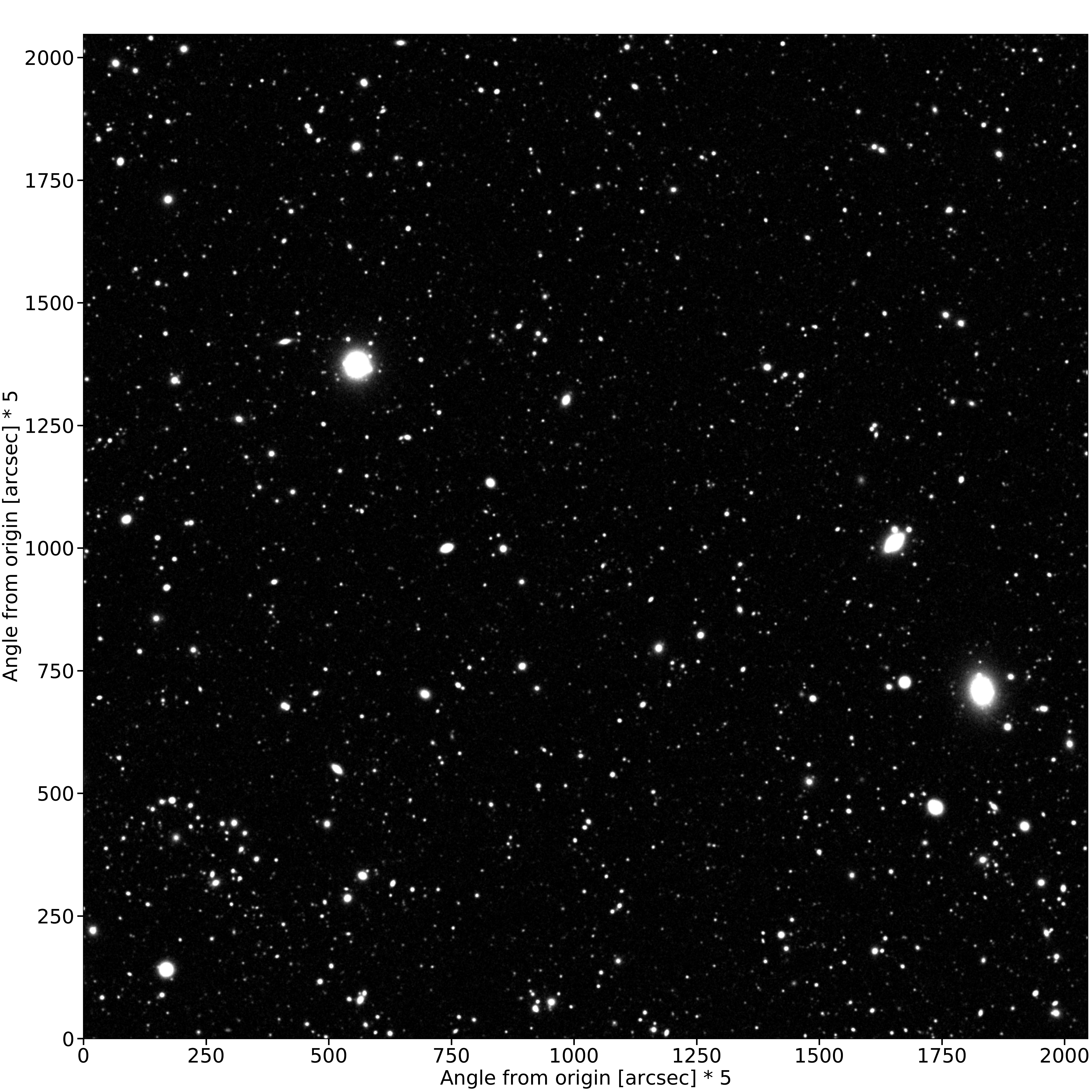}
\caption{Simulated \textit{i}-band coadd image, representing 100 exposures. To highlight as many galaxies as possible, the average sky background (estimated by taking the median pixel intensity) has been subtracted out, and the brightness has been scaled so that residual sky noise is visible as a background ``snow.'' \label{fig:scene}}
\end{figure}

Once raw pixel intensities are obtained for a given exposure, they are shifted by a fraction of a pixel, linearly interpolating their values, in such a way as to invert the effect of the random dither. Because the interpolated pixel value is not always well-defined for pixels at the extreme edge of an image, we ignore the outermost 1 pixel wide band around each image in everything that follows. For each image location, a total of 100 \textit{i}-band exposures, representing approximately half of the full LSST \textit{i}-band depth \citep{2019ApJ.873.2}, were simulated and finally added together into ``coadds.'' An example is shown in Figure~\ref{fig:scene}. The number of exposures roughly matches the depth reached in the DESC DC2 simulated survey. The choice of \textit{i}-band is motivated by the relatively high signal-to-noise ratio of objects in this band. This corresponds to the placement of \textit{i}-band at the top of the ``priority order'' the LSST Science Pipelines use for multi-band footprint processing \citep{2018PASJ.70.SP1.a}. \citet{2021_MuyGPs_StarGalaxy} found that the same band is most informative for training a Gaussian process classifier to distinguish stars from galaxies, using a method very similar to the one developed here. In section~\ref{sec:conclusions} we remark on a potential strategy for incorporating other filter bands. Each coadd takes a few hours to simulate on a commodity laptop. The image width of 2048 pixels is the largest power of 2 for which the simulation would fit in RAM.

The full DESC DC2 simulation includes several features that are of indirect relevance to the present study, such as Milky Way stars, optical reddening from Milky Way dust, Type Ia supernovae, and certain technical artifacts of the camera instrumentation and pointing. While these must necessarily be accounted for in a real survey, for this initial investigation we have elected to omit them for now and focus our attention on the distribution of galaxy properties alone. This simplifies our image processing requirements by allowing us to omit certain steps such as star--galaxy classification, corresponding to the assumption that an earlier stage in the pipeline has already done this and subtracted the stars (with idealized efficacy). This similarly streamlines the scope of the present analysis. The impact of, for example, stars misclassified as galaxies is expected to be the subject of future work.

\section{Data Preprocessing}\label{sec:preprocessing}
\begin{figure}
\plotone{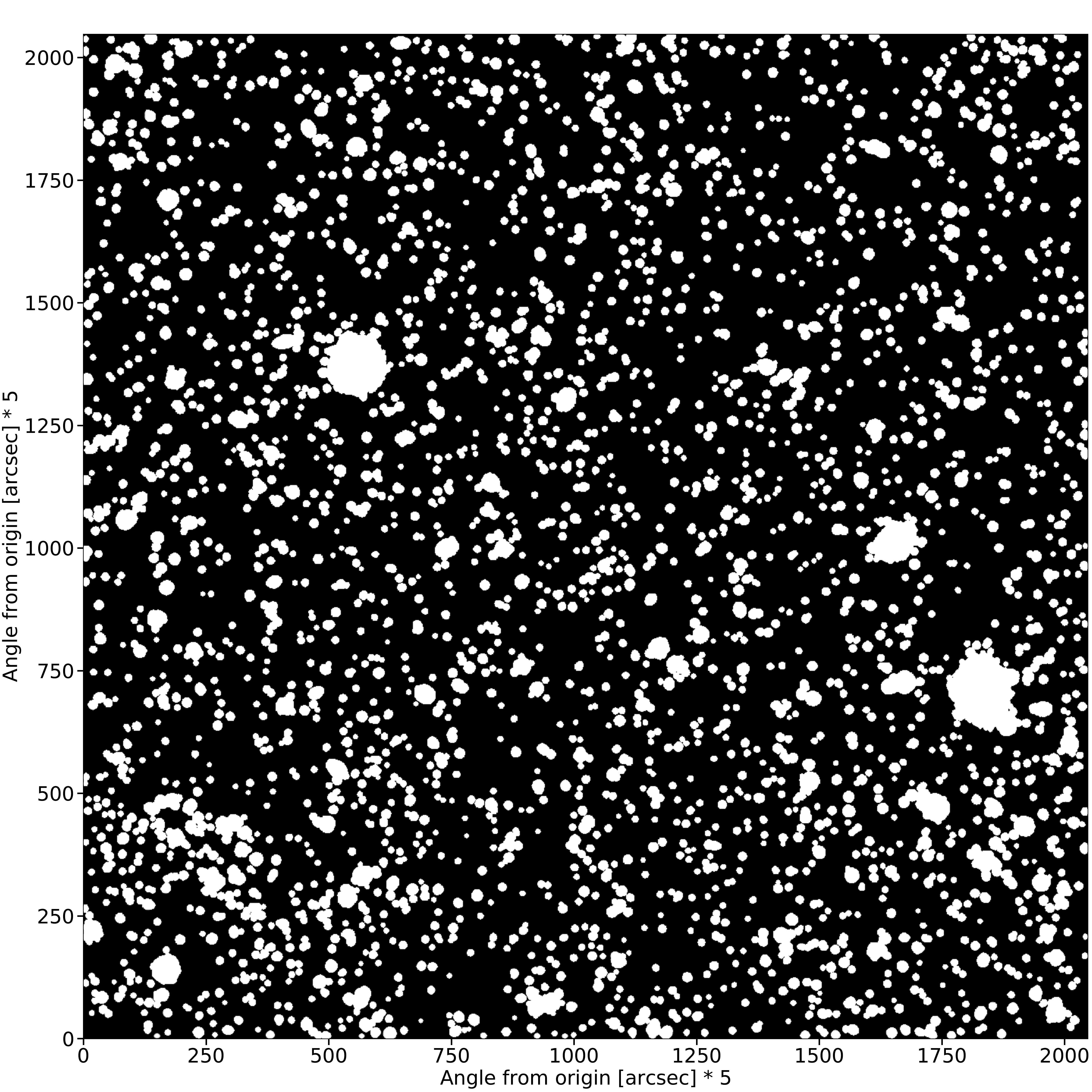}
\caption{Footprints produced by the current version of the LSST Science Pipelines, corresponding to the coadd shown in Figure~\ref{fig:scene}. \label{fig:footprints}}
\end{figure}

Certain preparatory steps, described in this section, must be performed on the coadds before their data are in a suitable format for the models investigated below. First, image regions containing galaxies must be identified with at least some preliminary, rough degree of precision. This is done in the HSC software pipeline \citep{2018PASJ.70.SP1.a}, and in the current version of the LSST Science Pipelines, by constructing ``footprints'' consisting of contiguous patches of high-intensity pixels. An initial set of preliminary footprints are first constructed using a relatively high intensity threshold. For simplicity of analysis, it is preferable if no candidate galaxies proposed at this stage are ``false'' detections arising purely from sky noise. Therefore, a pixel is only included in a preliminary footprint if its point-source signal-to-noise ratio (as defined in \citet{2018PASJ.70.SP1.a}) exceeds 5. Each footprint is then expanded to include all pixels up to 3 or 4 pixels away from every preliminary pixel, corresponding to roughly 2.4 times the radius of the PSF. Some expanded footprints may overlap, so any overlapping cases are finally merged into single footprints. Examples of footprints produced by the LSST Science Pipelines are shown in Figure~\ref{fig:footprints}.

The performance of a blend classification algorithm on footprints will naturally depend on the definition of a footprint. For this study we have compared the performance of classification algorithms on footprints constructed in three different ways. The first uses the current version of the LSST Science Pipelines, and the second is a custom reimplementation that uses median pixel intensity as its background estimation strategy, but otherwise tries to emulate the LSST Science Pipelines footprints as closely as possible. We found that a signal-to-noise threshold of 5.7, instead of 5.0, had to be used in the custom implementation for its footprint set to most closely resemble the Pipelines' in terms of number of total footprints and overall shape and size of footprints. In the figures that follow, these two footprint types are labeled ``Pipelines'' and ``Custom,'' respectively. The third footprint type, ``Variant,'' is a more significant variation from the Pipelines obtained by setting the custom implementation's threshold to 5.0 while also restricting the footprint expansion step to just pick up the immediate neighbors around each the of preliminary footprint pixels (corresponding to roughly 1.0 times the radius of the PSF). The specific footprint construction scheme to be adopted in the LSST Pipelines is open to future revision, and this variant strategy allows us to illustrate the impact on model performance of a relatively significant but still scientifically interesting variation in the footprint construction methodology.

For the purpose of classification, we define a footprint to be unblended if it contains the center of exactly one galaxy above a minimal brightness threshold. A footprint is blended if it contains the centers of two or more galaxies. A galaxy must have a point-source signal-to-noise ratio greater than 5 to be counted for this purpose, corresponding to an \textit{i}-band magnitude of 26.6. Our simulated image set contains a total of 134,493 such galaxies. A galaxy's center is defined to be the gnomonic projection onto the image plane, for a given simulated telescope pointing, of its lensed R.A.--decl.\ position listed in the Truth Match catalog. Changing the footprint definition affects the fraction of footprints that are blended vs.\ unblended, as well as the total number of footprints and the specific image patches made into footprints.  Table~\ref{tab:dataset_composition} provides a summary of dataset statistics for the three different footprint definitions. Occasionally, a footprint does not contain any galaxies. These ``empty'' footprints often correspond to the edges of actual galaxies whose centers lie just outside the image border, but they can also arise from galaxies slightly dimmer than the cutoff magnitude that happen to overlap with a sky noise over-fluctuation or the diffuse fringes of a large galaxy. For this initial study, we ignore empty footprints in everything that follows, since these comprise 1\% or less of all footprint sets considered here.

To compute the galaxy signal-to-noise ratio, we use the same equation for the point-source signal-to-noise ratio for pixels given in \citet{2018PASJ.70.SP1.a}, but with the observed pixel intensities replaced with their expected intensities for a point source with a given location and magnitude. The same (pixel-level) equation is used to select the initial pixels for each of the different footprint types, prior to expansion. However, the LSST Science Pipelines ultimately estimate a different signal-to-noise value for each pixel, compared to the independent estimate we use to construct the Custom and Variant footprints. Differences between the two estimates essentially amount to differences in the assumed sky background level, background noise, and PSF characteristics --- the Pipelines estimate these values based on various image characteristics, whereas in our independent estimate we use the true mean sky background, sky noise, and PSF widths used in the image simulation. For our galaxy signal-to-noise estimate, which establishes the magnitude threshold of 26.6, we use the same true values. If the Pipeline-estimated values were used instead, we anticipate that the corresponding galaxy signal-to-noise ratios would decrease, resulting in a lower magnitude (higher brightness) cutoff. In comparison, using the higher value of 26.6 as the cutoff somewhat increases the probability that any given footprint is blended, decreases the fraction of all galaxies above cutoff that are contained in any footprint, and decreases the number of empty footprints.

For all classification methods other than peak finding, each footprint is then assigned to a ``cutout'' initially consisting of a square subarray of image pixels, of some fixed size, centered on the footprint. The value of every pixel in the cutout that is not part of the footprint is set to zero. The inputs to the Gaussian process model are one-dimensional arrays of numbers, also called vectors, and so at this stage the two-dimensional cutout is ``flattened'' by rearranging its pixel values into a single row. All values in the vector are then normalized to have absolute values no greater than 1. Finally, reducing the dimensionality of the input vectors using PCA has been shown to improve the speed and accuracy of Gaussian process models used for star--galaxy classification \citep{2021_MuyGPs_StarGalaxy}, and it was similarly found to improve performance in the present setting. In PCA, some set of orthonormal vectors --- principal components --- is identified that maximizes the componentwise variance of the training data when projected onto that set. All training and testing data are finally projected onto that set of principal components. Reducing the dimensionality of the dataset amounts to choosing a number of principal components that is lower than the length of the original data vectors.

The cutout width and PCA-embedding dimension may be thought of as model hyperparameters, as their choices affect the ultimate classification performance. In our tests we found that cutout widths of at least 23 pixels together with an embedding dimension between 7 and 10 resulted in the best balanced accuracy (defined below) seen in \textit{k}-fold cross-validation on the training data. If these hyperparameters are set accordingly, together with the Gaussian process hyperparameter values discussed below, then other hyperparameters (such as the specific cutout centering strategy or the normalization strategy) have comparatively negligible impact. The ``optimal'' choice of hyperparameters ultimately depends on the specific science objective(s) under investigation, with classification accuracy being just one of many possible metrics to consider. The performance of these models with respect to specific scientific objectives, and the corresponding hyperparameter choices that optimize those objectives, are expected to be the subject of future work.
\begin{deluxetable*}{lDDD}
\tablecaption{Composition of footprint sets made according to three different prescriptions, for simulated LSST images as described in section~\ref{sec:imagesim}. The columns correspond to the three footprint prescriptions in the order described in section~\ref{sec:preprocessing}. The ``Galaxies'' row specifies percentages of the total number of galaxies with a point-source signal-to-noise ratio exceeding 5. The other rows specify percentages of the total number of footprints made according to a given prescription. Numbers in parentheses indicate statistical uncertainties in the final digits of the preceding value, arising from the finite size of the footprint samples.\label{tab:dataset_composition}}
\tablehead{ & \multicolumn{6}{c}{Footprint Type} \\
 & \twocolhead{Pipeline} & \twocolhead{Custom} & \twocolhead{Variant}}
\decimals
\startdata
Galaxies contained in \textit{i}-band footprints (\%)    & 66.01(13) & 64.83(13) & 69.67(13) \\
Unblended footprints (\%)                       & 61.74(22) & 63.86(22) & 73.08(18) \\
Blended footprints (\%)                         & 38.01(22) & 35.04(22) & 25.58(17) \\
Empty footprints (\%)                           & 0.25(2) & 1.10(5) & 1.33(5)
\enddata
\end{deluxetable*}

\section{Models}
In this section we introduce each of the models we analyze for the task of blend classification. These include the peak-finding method (section~\ref{sec:peak_finding}) along with several probabilistic models, including Gaussian process (section~\ref{sec:gp_model}), CNN (section~\ref{sec:cnn_model}), and logistic regression (section~\ref{sec:lr_model}) models.
\subsection{Peak Finding}\label{sec:peak_finding}
One straightforward method of estimating the galaxy distribution in an image is to assert that galaxies are in one-to-one correspondence with local brightness peaks. In the present study (corresponding to the HSC/LSST pipeline implementation) we find peaks by convolving the image with a Gaussian approximation of its PSF, and then identifying pixels in the convolved image that (a) lie in a footprint and (b) have intensities at least as large as all eight of their immediate neighbors.\footnote{If two immediately neighboring pixels, having equal PSF-convolved intensities, could be identified as peaks in this way, one of these (chosen arbitrarily) is ignored.} This method has several immediate advantages: it requires no pre-training, peak positions can be computed rapidly, and these simultaneously estimate both the number and the centroid locations of galaxies in an image. One disadvantage is that it is not immediately clear how probable these best-guess values, or other choices of values, might actually be. To probe this issue, we studied the reliability of peak finding as a blend classifier on the three types of footprints defined above in section~\ref{sec:preprocessing}. Table~\ref{tab:peakfinding_acc} summarizes its overall accuracy.
\begin{deluxetable*}{lDDD}
\tablecaption{Accuracy of peak finding for footprint blend classification. The columns correspond to the three footprint prescriptions in the order described in section~\ref{sec:preprocessing}. The third row is the mean of the first two rows. Numbers in parentheses indicate statistical uncertainties in the final digits of the preceding value, due to the finite size of the image set.\label{tab:peakfinding_acc}}
\tablehead{ & \multicolumn{6}{c}{Footprint Type} \\
 & \twocolhead{Pipeline} & \twocolhead{Custom} & \twocolhead{Variant}}
\decimals
\startdata
Unblended footprints with 1 peak (\%)    & 75.44(25) & 98.27(07) & 99.70(3) \\
Blended footprints with ${>}1$ peak (\%) & 78.85(30) & 71.43(35) & 52.61(39) \\
Balanced accuracy (\%)                   & 77.14(20) & 84.85(18) & 76.15(20)
\enddata
\end{deluxetable*}

\subsection{Gaussian Process Model}\label{sec:gp_model}
Gaussian process regression has been used to classify stars vs.\ galaxies in simulated LSST images \citep{2021_MuyGPs_StarGalaxy}, and a substantially similar model is implemented here for galaxy blend classification. 
A Gaussian process is a collection of random variables for which any finite subset is Gaussian-distributed. 
In this case, the random variables are real numbers specifying the ``blendedness'' of a footprint, one for every possible real-valued vector with dimensionality equal to the PCA-embedding dimension used in footprint preprocessing. We begin by splitting our dataset into an equal number of training and testing images, and assume a Bayesian prior on the blendedness values of all examples in the dataset, given as a function of the PCA vectors. Specifically, we assert that this prior is a Gaussian distribution, with a mean of 0 for each variable, and a covariance matrix composed of four kernel submatrices denoted $K_\mathbf{ff}$, $K_\mathbf{f*}$, $K_\mathbf{*f}$, and $K_\mathbf{**}$. The entries in each kernel submatrix are defined via a kernel function $k$ as follows:
\begin{eqnarray}
(K_\mathbf{ff})_{i,j} \equiv k(x_{i}^{\text{train}}, x_{j}^{\text{train}}) \nonumber\\
(K_\mathbf{f*})_{i,j} \equiv k(x_{i}^{\text{train}}, x_{j}^{\text{test}}) \\
(K_\mathbf{*f})_{i,j} \equiv (K_\mathbf{f*})_{j,i} \nonumber\\
(K_\mathbf{**})_{i,j} \equiv k(x_{i}^{\text{test}}, x_{j}^{\text{test}}) \nonumber
\end{eqnarray}
where $x_i$ is the PCA vector for example $i$.

We assign ``hard'' blendedness values $\mathbf{y}$ to all training examples, such that for each example $i$, $y_i = +1$ if it is blended and $-1$ if it is unblended. To account for random noise, we assume that $\mathbf{y}$ is the sum of two random variables, one of which is a vector-valued function $\mathbf{f}$ of the training examples' PCA vectors, and the other corresponding to a random perturbation, uncorrelated between examples, with homoscedastic variance $\tau^2$. The unseen blendedness values of the test examples are represented by the vector $\mathbf{f_*}$. The full prior distribution is then
\begin{equation}
\begin{bmatrix}
\mathbf{y} \\ 
\mathbf{f_*}
\end{bmatrix} = \mathcal{N} \left (0, \sigma^2 \begin{bmatrix}
K_\mathbf{ff} + \tau^2 I_n & K_\mathbf{f*} \\ 
K_\mathbf{f*} & K_\mathbf{**}
\end{bmatrix} \right )
\end{equation}
where $I_n$ is the identity matrix of rank $n$ equal to the number of training examples, and $\sigma$ is a scale factor that sets the overall variance in the prior. In order to justifiably assume a prior mean of 0 everywhere, we ensure that the training data contain equal numbers of blended and unblended examples, by keeping only as many unblended footprints (selected at random) as there are blended footprints. To account for this random element in dataset construction, we repeat all model evaluations 100 times, making a new random dataset choice each time. The accuracy values reported here are the averages over those 100 trials.

Given the known values of $\mathbf{y}$, we then analytically compute the posterior joint distribution of $\mathbf{f_*}$ via
\begin{equation}
\mathbf{f_*} | X^\text{train}, X_*^\text{test}, \mathbf{y} \sim \mathcal{N} ( \mathbf{\bar{f}_{*}}, \sigma^2 C )
\end{equation}
where
\begin{eqnarray}
\mathbf{\bar{f}_{*}} \equiv K_\mathbf{*f} ( K_\mathbf{ff} + \tau^2 I_n )^{-1} \mathbf{y}, \\
C \equiv K_\mathbf{**} - K_\mathbf{*f} ( K_\mathbf{ff} + \tau^2 I_n )^{-1} K_\mathbf{f*}.
\end{eqnarray}
Because the posterior, like the prior, is a Gaussian distribution, most of its probability mass lies on the same side of 0 as $\mathbf{\bar{f}_{*}}$. The most probable sign (either $+1$ or $-1$) of the blendedness of a test example $i$ is therefore the sign of $\bar{f}_{\mathbf{*}i}$. As such, we classify a test example $i$ as blended or unblended according to this sign.

The kernel $k$ should be a symmetric, positive semidefinite function of its inputs \citep{rasmussenandwilliams}, but otherwise its specific functional form is a tunable hyperparameter. In tests we found that a Matérn kernel of the form\footnote{This is the same form for the Matérn kernel used in e.g.\ \citet{rasmussenandwilliams}. In the same text it is remarked that $k$ can be multiplied by $\sigma^2$ to set any arbitrary level of overall variance. Our convention here for the place of $\sigma$, corresponding to \citet{2021_MuyGPs_StarGalaxy}, is effectively equivalent, and represents how the computations are actually performed in our analysis.}
\begin{equation}
k(x, x') = \frac{2^{1-\nu}}{\Gamma(\nu)} \left ( \frac{\sqrt{2\nu}||x - x'||_2}{l} \right )^\nu K_\nu \left ( \frac{\sqrt{2\nu}||x - x'||_2}{l} \right )
\end{equation}
resulted in the best balanced accuracy seen in \textit{k}-fold cross-validation on the training data. Here $K_\nu$ (with a single subscript) refers to a modified Bessel function of the second kind, and $||x - x'||_2$ is the Euclidean distance between the input vectors $x$ and $x'$. This kernel has two tunable values $\nu$ and $l$, corresponding to the ``smoothness'' and ``length scale'' of correlations between blendedness values at different points in the PCA embedding space. It equals 1 whenever both of its inputs are equal and approaches zero for infinitely distant inputs, making it a measure of similarity between inputs.

Aside from computing the kernel submatrices and inverting $( K_\mathbf{ff} + \tau^2 I_n )$, ``training'' the Gaussian process model amounts to tuning the various hyperparameters. For this study, we have done this by maximizing the balanced accuracy seen in 400-fold cross-validation on the training data, where balanced accuracy is defined to be the mean of two values: the fraction of blended footprints classified as blended, and the fraction of unblended footprints classified as unblended. Some hyperparameters involve steps of the data preprocessing pipeline, the most important of which are the PCA-embedding dimension and the cutout width, as remarked in section~\ref{sec:preprocessing}. Others belong to the Gaussian process model itself, including the choice of kernel function, the free parameters of the kernel (in this case, $\nu$ and $l$), and the uncorrelated noise level $\tau$. In our tests, values of $\nu$ greater than 1, values of $l$ between 10 and 100, and values of $\tau$ between $1e{-6}$ and $1e{-4}$ gave the best performance, with similar results seen all throughout those ranges.

The specific hyperparameter values chosen for model comparison are listed in Table~\ref{tab:hyperparameter_values}. Rapid hyperparameter tuning was enabled by MuyGPs \citep{2021_MuyGPs}, a novel code for efficient Gaussian process inference. One distinguishing feature of MuyGPs is the use of only some fixed number of nearest neighbors, in feature space, for posterior inference. Here we use 50 nearest neighbors, and ``nearest'' here means least Euclidean distance in PCA embedding space. MuyGPs was also used to identify the leading PCA components of the training data. Initial values of $\nu$, $l$, and $\tau$ were selected by running MuyGPs' automatic hyperparameter optimization routine several times and selecting the most typical results for each. With those values fixed, the PCA-embedding dimension and cutout width were then optimized using a grid search with successively finer grids ultimately converging on the values listed in Table~\ref{tab:hyperparameter_values}. The Gaussian-process-specific parameters $\nu$, $l$, and $\tau$ were then optimized similarly, using fixed values of the PCA dimension and cutout width. A similarly optimized model with a squared-exponential radial basis function kernel instead of Matérn did not perform as well. No further improvement was seen in a second round of grid search over PCA dimension and cutout width. Various alternative cutout normalization, background subtraction, cutout centering, and nearest-neighbor schemes were tested, for a small range of hyperparameter values around their table values, but no further improvement was seen. The overall prior width $\sigma$ only enters into the posterior through the covariance, and so does not directly impact the model accuracy (which is only a function of the posterior mean); $\sigma$ is tuned using a separate procedure described in section~\ref{sec:uq}. Table~\ref{tab:gp_acc} summarizes the overall accuracy of this model on each type of footprint.
\begin{deluxetable*}{ll}
\tablecaption{Hyperparameter choices used for all Gaussian process performance metrics reported in this study.\label{tab:hyperparameter_values}}
\tablehead{\colhead{Hyperparameter} & \colhead{Setting}}
\startdata
Cutout width & 23 pixels \\
Cutout center location & Footprint intensity barycenter \\
Cutout background subtraction & None \\
Cutout normalization & Divide each pixel value by max value in training set \\
PCA-embedding dimension & 8 \\
GP kernel & Matérn \\
Matérn smoothness ($\nu$) & 10 \\
Matérn length scale ($l$) & 20 \\
Noise level ($\tau$) & $3e{-6}$ \\
Number of nearest neighbors & 50 \\
Nearest neighbor metric & Euclidean
\enddata
\end{deluxetable*}
\begin{deluxetable*}{lDDD}
\tablecaption{Accuracy of Gaussian process blend classification. The columns correspond to the three footprint prescriptions in the order described in section~\ref{sec:preprocessing}. The third row is the mean of the first two rows. Numbers in parentheses indicate statistical uncertainties in the final digits of the preceding value, due to the finite size of the test set. Numbers in square brackets indicate the root-mean-square variation in the final digits of the preceding values due to variations in the training set assembly.\label{tab:gp_acc}}
\tablehead{ & \multicolumn{6}{c}{Footprint Type} \\
 & \twocolhead{Pipeline} & \twocolhead{Custom} & \twocolhead{Variant}}
\decimals
\startdata
Unblended footprints classified as unblended (\%) & 94.20(19)[14] & 94.17(19)[19] & 91.58(18)[16] \\
Blended footprints classified as blended (\%)     & 80.04(41)[13] & 82.41(40)[19] & 81.09(43)[22] \\
Balanced accuracy (\%)                            & 87.12(23)[07] & 88.29(22)[06] & 86.34(23)[06]
\enddata
\end{deluxetable*}

\subsection{CNN Model}\label{sec:cnn_model}
The CNN model comprises several sequential layers, listed in Table~\ref{tab:cnn_layers}. Model input consists of the same 23x23 single-channel footprint cutouts used for the Gaussian process model, for which the pixels not belonging to the footprint have all been set to zero. These otherwise raw inputs are preprocessed in two stages: first a nonlinear ``stretch'' is applied to the pixel values by replacing them with their base-10 logarithm,\footnote{To avoid taking the log of 0, $10^{-8}$ is added to each pixel value beforehand.} and then the resulting pixel values are min--max rescaled. The first neural network layer, consisting of 128 two-dimensional convolutional filters each with a 3x3 receptive field, takes these preprocessed cutouts as input. This layer's weighted inputs pass through a rectified linear unit (ReLU) activation function followed by a max pooling layer that reduces the filtered image width by half. This is followed by a second convolutional layer, the same as the first but with 64 filters, and a second max pooling layer that again halves the filtered image width. The two-dimensional output of this second pooling layer is flattened to a one-dimensional array and input to a dense hidden layer with 800 units, then one with 400 units, and finally one with 200 units, each with a ReLU activation function. Dropout layers are interspersed between the 800-400 layers and the 400-200 layers, both with a dropout probability of 20\%. The final layer of the network is a softmax layer with two outputs constrained to lie between 0 and 1. One of the two softmax outputs is arbitrarily selected to be a blended score and the other as an unblended score, and a given footprint is classified as blended just when the blended score is higher than the unblended score. The network is trained to minimize the cross-entropy loss on the training data, using the Adam optimizer with batches of 200 randomly selected training footprints in each iteration. The balanced accuracy on held-out validation data (10\% of the total training data) stopped improving after 15 full passes through the training data, at which point training was terminated. The model architecture, data preprocessing procedure, and training procedure are identical to those used by \citet{2021_MuyGPs_StarGalaxy} for classification of stars vs. galaxies. Table~\ref{tab:cnn_acc} shows the overall accuracy of the CNN model on test data.
\begin{deluxetable*}{ll}
\tablecaption{The data preprocessing steps and neural network layers of the CNN model used in this analysis, listed in order from earliest at the top to latest at the bottom. All Convolution and Dense layers use a ReLU activation function.\label{tab:cnn_layers}}
\tablehead{\colhead{Layer} & \colhead{Description}}
\startdata
Image [input] & 23x23, 1 channel \\
Nonlinear stretch & $\log_{10}(\text{pixel value} + 10^{-8})$ \\
Min-max rescale & $(\text{pixel value} - \text{min}) / (\text{max}-\text{min})$ \\
Convolution & 128 3x3 filters, stride 1 \\
Max Pool & 2x2 pool size, stride 2 \\
Convolution & 64 3x3 filters, stride 1 \\
Max Pool & 2x2 pool size, stride 2 \\
Dense & 800 units \\
Dropout & 0.2 probability \\
Dense & 400 units \\
Dropout & 0.2 probability \\
Dense & 200 units \\
Softmax [output] & 2 classes
\enddata
\end{deluxetable*}
\begin{deluxetable*}{lDDD}
\tablecaption{Accuracy of convolutional neural network blend classification. The columns correspond to the three footprint prescriptions in the order described in section~\ref{sec:preprocessing}. The third row is the mean of the first two rows. Numbers in parentheses indicate statistical uncertainties in the final digits of the preceding value, due to the finite size of the test set. Numbers in square brackets indicate the root-mean-square variation in the final digits of the preceding values due to variations in the training set assembly.\label{tab:cnn_acc}}
\tablehead{ & \multicolumn{6}{c}{Footprint Type} \\
 & \twocolhead{Pipeline} & \twocolhead{Custom} & \twocolhead{Variant}}
\decimals
\startdata
Unblended footprints classified as unblended (\%) & 97.67(12)[78] & 97.62(12)[55] & 97.87(09)[56] \\
Blended footprints classified as blended (\%)     & 75.27(44)[183] & 77.52(44)[167] & 66.34(51)[323] \\
Balanced accuracy (\%)                            & 86.47(12)[99] & 87.57(12)[88] & 82.10(09)[164]
\enddata
\end{deluxetable*}

\subsection{Logistic Regression Model}\label{sec:lr_model}
To provide a baseline for comparison among the probabilistic models, we finally consider a simple logistic regression classifier. The logistic regression model was optimized by minimizing the cross-entropy loss, as defined in section~\ref{sec:uq}, on the training set. It takes as input the same preprocessed, PCA-embedded data vectors as the Gaussian process model. Neighboring values of the PCA-embedding dimension and cutout width did not significantly improve the logistic regression model's balanced accuracy in 50-fold cross-validation, so for simplicity the same values were used for this model as for the Gaussian process model. In order to potentially improve its generalizability, an additional term proportional to the sum of the squared model weights was added to the cross-entropy loss during training. The constant of proportionality for this extra term was optimized with respect to balanced accuracy on five-fold cross-validation within the training set, which ran as an ``inner'' loop inside the ``outer'' 50-fold cross-validation used to optimize the other hyperparameters. Table~\ref{tab:lr_acc} shows the overall accuracy of the logistic regression model on test data.
\begin{deluxetable*}{lDDD}
\tablecaption{Accuracy of logistic regression blend classification. The columns correspond to the three footprint prescriptions in the order described in section~\ref{sec:preprocessing}. The third row is the mean of the first two rows. Numbers in parentheses indicate statistical uncertainties in the final digits of the preceding value, due to the finite size of the test set. Numbers in square brackets indicate the root-mean-square variation in the final digits of the preceding values due to variations in the training set assembly.\label{tab:lr_acc}}
\tablehead{ & \multicolumn{6}{c}{Footprint Type} \\
 & \twocolhead{Pipeline} & \twocolhead{Custom} & \twocolhead{Variant}}
\decimals
\startdata
Unblended footprints classified as unblended (\%) & 89.51(25)[60] & 86.90(27)[16] & 82.56(25)[74] \\
Blended footprints classified as blended (\%)     & 71.55(47)[53] & 77.97(44)[25] & 72.91(48)[78] \\
Balanced accuracy (\%)                            & 80.53(26)[08] & 82.43(26)[10] & 77.74(27)[09]
\enddata
\end{deluxetable*}
\section{Model Comparison}\label{sec:comparison}
\begin{figure}
\includegraphics[width=\textwidth]{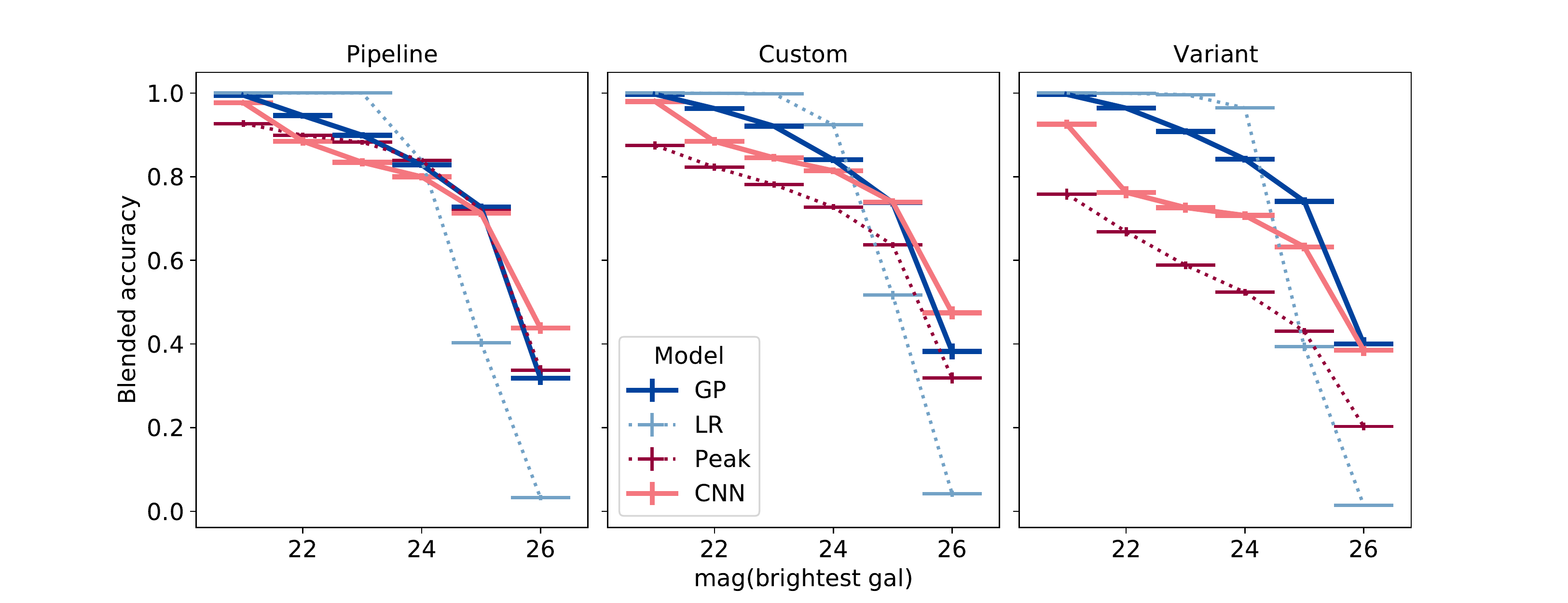}
\caption{Fraction of blended pipeline footprints in the test set classified as blended by each model, plotted as a function of the magnitude of the brightest galaxy in each footprint. The vertical error bars indicate statistical uncertainty due to the finite size of the test set. The horizontal error bars span the entire magnitude range corresponding to each point.\label{fig:blend_acc_comparison}}
\end{figure}

\begin{figure}
\includegraphics[width=\textwidth]{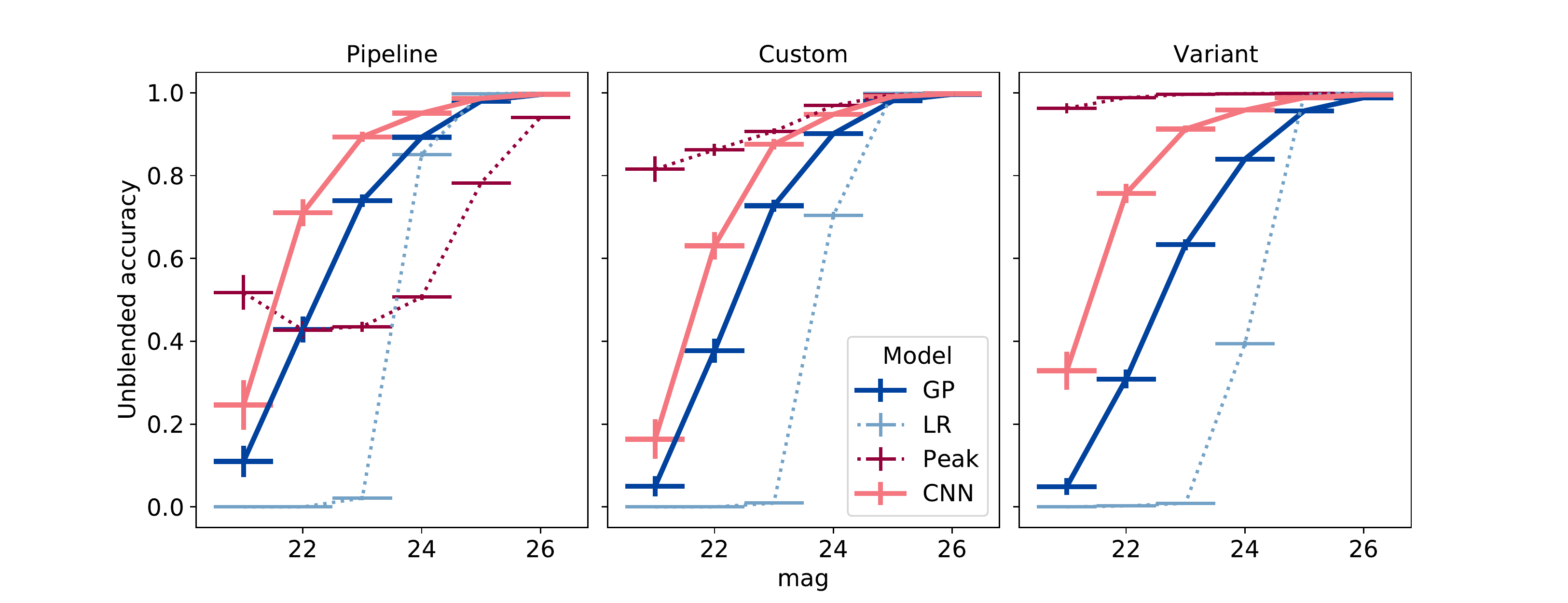}
\caption{Fraction of unblended pipeline footprints in the test set classified as unblended by each model, plotted as a function of the magnitude of the only galaxy in each footprint. The vertical error bars indicate statistical uncertainty due to the finite size of the test set. The horizontal error bars span the entire magnitude range corresponding to each point.\label{fig:unblend_acc_comparison}}
\end{figure}

Figures~\ref{fig:blend_acc_comparison} and \ref{fig:unblend_acc_comparison} compare the classification accuracy as a function of magnitude among several models. The differences in peak finding accuracies between the Pipeline and Custom footprints are striking given that the pixel intensity threshold of the Custom footprints was tuned so that the resulting footprint set resembled the Pipelines footprints as closely as possible in both number and shape. These results indicate that the peak structure of footprints depends quite sensitively on the specifics of background and noise estimation. Variant footprints are expanded less than the other footprint types, leading them to be substantially smaller in general and hence substantially less likely to contain more than one peak, resulting in a distinct bias in favor of classifying everything as unblended. This connection between footprint size and classification bias is recapitulated in the trend of each accuracy plot: for all models and all footprint types, these plots exhibit monotonically increasing biases in favor of unblended classifications as magnitude increases,\footnote{Except for a slight dip in the Pipeline plot of unblended accuracy for peak finding, though the size of the vertical error bars indicates this feature has relatively low statistical significance.} and larger magnitudes generally indicate smaller footprint sizes.
On Pipeline footprints, the Gaussian process model performs just as well (within statistical uncertainties) or better than the peak-finding method in almost all contexts. The one exception is unblended pipeline footprints at low magnitudes, for which peak finding more reliably classifies them as unblended. All of the probabilistic models, in particular the Gaussian process model (for a fixed choice of hyperparameters), are substantially less sensitive than the peak-finding method to the specific choice of footprint definition. Furthermore, while the specific Gaussian process hyperparameters listed in Table~\ref{tab:hyperparameter_values} were optimized with respect to the Custom footprint definition, these same values give cross-validation accuracies that are within statistical uncertainty of the best values observed for every other type of footprint, justifying the use of these same hyperparameters for all footprint types.

As shown in Figures~\ref{fig:blend_acc_comparison} and \ref{fig:unblend_acc_comparison}, the logistic regression model evidently latches on very strongly to the overall brightness of a footprint, making its classification prediction predominantly on that basis alone. It nearly always predicts that the very brightest footprints are blended, and generally predicts that the dimmest footprints are unblended, regardless of the arrangement of light within a footprint. As remarked above, the other models also exhibit biases along these lines, though in a more subdued fashion. Some of this may be due to the fact that, while the training data were evenly split between blended and unblended footprints overall, the split within any specific magnitude range was not guaranteed to be even, and in practice was often significantly imbalanced. In particular, in the lowest magnitude bin shown in the figures, more than four fifths of footprints in the training data are blended. Evening out the training set class representation at different magnitude levels might improve the logistic regression and Gaussian process model predictions. However, this also decreases the amount of available training data, requiring a larger dataset to achieve the same level of statistical precision. Furthermore, the peak-finding method still exhibits some of the same bias even though it is not trained, suggesting that some of this bias may be difficult to alleviate even with a more thoroughly balanced training set.

The CNN classifier generally has a higher unblended accuracy and a lower blended accuracy than the Gaussian process classifier across all magnitudes and footprint types. The two models generally seem to make different kinds of mistakes, leading us to conjecture that in some average sense, many of their errors might ``cancel out,'' and that some combination of the two models might be a substantially more accurate classifier than either model in isolation. For now we note that the Gaussian process classifier's accuracy is quite competitive with the CNN classifier's. Furthermore, the Gaussian process model is faster to train: in the time it takes to train a single CNN instantiation, the Gaussian process model can be re-solved multiple times with different input and model settings, allowing rapid hyperparameter tuning. The Gaussian process model also requires only about a fifth of the memory needed by the CNN, indicating that it may be significantly more parallelizable. However, the CNN model is faster at classifying any single example, and it benefits in particular from GPU acceleration. The MuyGPs Gaussian process implementation does not currently enable a similar form of GPU acceleration, though this is a topic of ongoing study. In comparison, peak finding is effectively a ``free'' operation with trivial requirements in both time and memory relative to the resources needed to construct footprints, and it never needs to be trained.

\section{Uncertainty Quantification}\label{sec:uq}
In addition to the predicted class labels, inferred from the sign of the posterior mean $\mathbf{\bar{f}_{*}}$, the Gaussian process model also supplies a quantitative estimate the confidence in its predictions. By integrating the posterior distribution of $f_{\mathbf{*}i}$ from 0 to infinity (i.e.\ over all positive values), we obtain the total posterior probability that footprint $i$ has a positive $f$-value, which we take to be the model-estimated probability that the footprint is blended. This integral depends on the value of $\sigma$, which does not directly affect the posterior mean and so could not be tuned, like the other hyperparameters, by optimizing the classification accuracy.

Instead, we choose a value of $\sigma$ that maximizes the model-assigned likelihood on validation examples. Specifically, and equivalently, we minimize the cross-entropy loss, defined as
\begin{equation}
L_{ce} = -\sum_{i} [y_{i} \log(p_{i}) + (1 - y_{i}) \log(1 - p_{i})]
\end{equation}
where $y_{i}$ is the true class label of footprint $i$, $p_{i}$ is the model-assigned posterior blend probability, and the sum is taken over all validation examples in all held-out folds used in 400-fold cross-validation on the training set. Since the posterior probability depends on the training examples used, and since the training set assembly includes an element of randomness, we tuned $\sigma$ based on the mean cross-entropy loss computed over 100 random training set initializations. Values of $\sigma$ were considered in successively narrowing grids, converging to minimum cross-entropy losses for each footprint type at the values listed in Table~\ref{tab:sigma_values}. These values are several orders of magnitude larger than the range of target mean prediction values ($\pm1$), indicating that the unscaled posterior variance $C$ is typically very small in comparison to this range. The smallness of $C$ in turn indicates that the typical distances between test footprints and their nearest training footprints, in PCA embedding space, are relatively small compared to the kernel length scale $l$. It would seem that even such relatively nearby points in feature space can routinely have different class labels, necessitating a large value of $\sigma$ to scale the small values of $C$ sufficiently high enough to cover an appropriate uncertainty range.
\begin{deluxetable*}{lD}
\tablecaption{Optimal $\sigma$ values for each footprint type.\label{tab:sigma_values}}
\tablehead{Footprint Type & \twocolhead{$\sigma$}}
\decimals
\startdata
Pipeline & 1500 \\
Custom & 1800 \\
Variant & 1600
\enddata
\end{deluxetable*}

The logistic regression classifier is similarly optimized by treating its outputs as class probability estimates, and adjusting its parameters via gradient descent so as to minimize the corresponding cross-entropy loss on the training set. Generally speaking, this ensures that probability estimates are as discriminating as possible while also preserving their reliability. Table~\ref{tab:mean_cross_entropy} lists the mean cross-entropy loss for each probabilistic model (i.e.\ every model other than peak finding) on the test set. From this we infer that the Gaussian process model will be more likely than the logistic regression model to at once classify test examples correctly, place a high degree of confidence in its correct predictions, and place a lower degree of confidence in incorrect predictions. The softmax value of the CNN model's two outputs is treated as a blend probability estimate, and like the logistic regression model, the CNN model parameters are tuned to minimize the cross-entropy loss on the training set. It exhibits the lowest test set cross-entropy loss among the models we consider here. One reason seems to be its frequent willingness to assign scores extremely close to 0 and 1. These scores are not always correct, but they are correct just often enough to result in a relatively low cross-entropy loss.
\begin{deluxetable*}{lDDD}
\tablecaption{Mean cross-entropy loss over all test footprints, for each probabilistic model and each footprint type. Numbers in parentheses indicate statistical uncertainties in the final digits of the preceding values, due to the finite size of the test set. Numbers in square brackets indicate the root-mean-square variation in the final digits of the preceding values, due to variations in the training set assembly.\label{tab:mean_cross_entropy}}
\tablehead{ & \multicolumn{6}{c}{Footprint Type} \\
\colhead{Model} & \twocolhead{Pipeline} & \twocolhead{Custom} & \twocolhead{Variant}}
\decimals
\startdata
Gaussian process    & 0.3235(38)[05] & 0.3006(37)[05] & 0.2959(29)[09] \\
Convolutional neural network & 0.3011(50)[81] & 0.2708(49)[76] & 0.2975(49)[112] \\
Logistic regression & 0.4384(37)[14] & 0.4033(35)[36] & 0.4742(26)[99]
\enddata
\end{deluxetable*}

\begin{figure}
\centering
\includegraphics[width=0.5\textwidth]{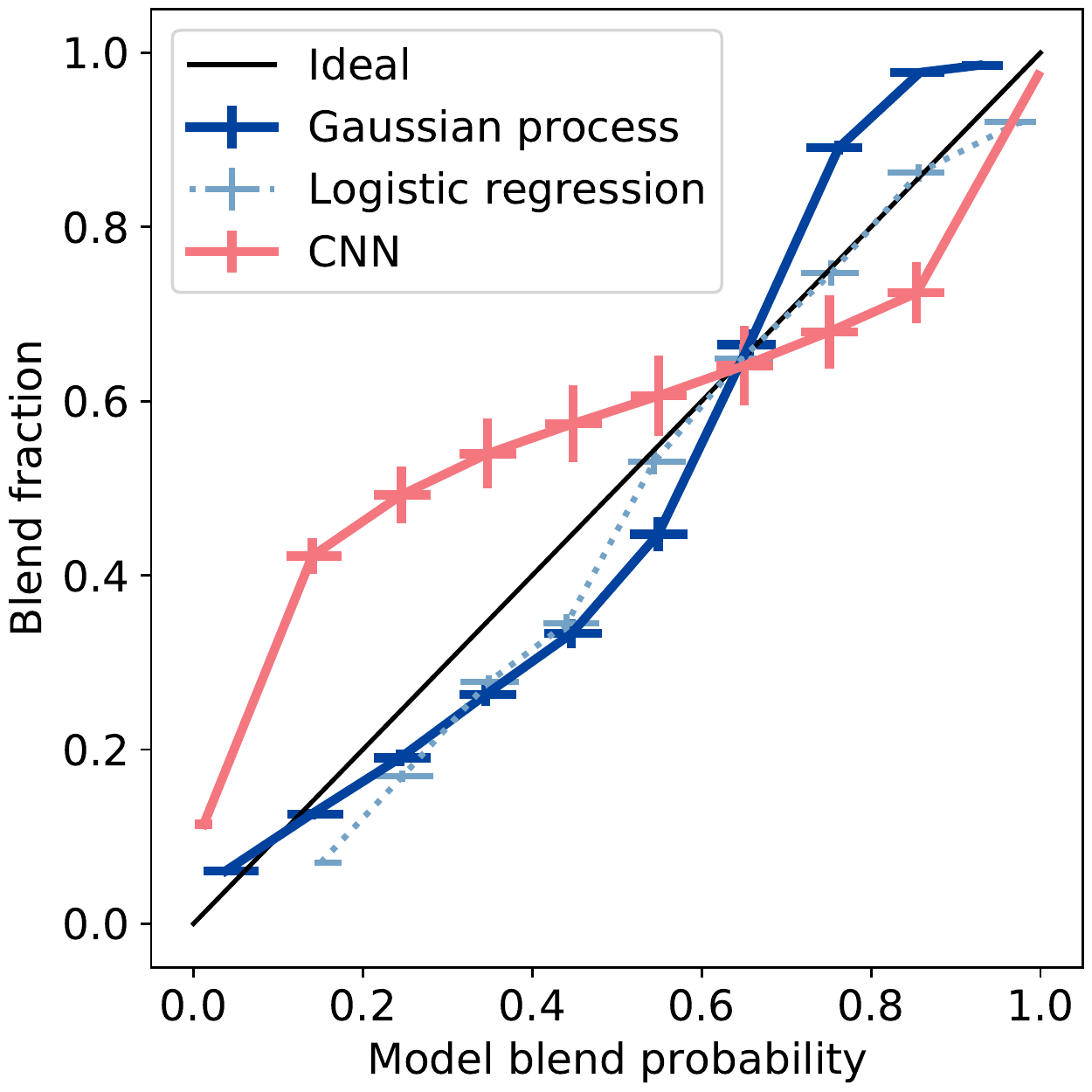}
\caption{Probabilistic calibration of each probabilistic model on Pipeline footprints, evaluated in 10 equally spaced horizontal bins from 0\% to 100\%. For each bin of model-assigned probability, the vertical axis specifies the fraction of pipeline footprints in the test set that are actually blended. The horizontal coordinate of each plotted point equals the median model-assigned probability among test footprints in a given horizontal bin; the horizontal error bars cover 68\% of test examples in each horizontal bin. The vertical error bars indicate statistical uncertainty due to the finite size of the test set. The black diagonal line indicates ``ideal'' performance, in the sense that for a model with a calibration point centered on that line, that model's estimated blend probability generally matches the fraction of pipeline footprints that are actually blended.\label{fig:calibration_comparison}}
\end{figure}

Intuitively, one expects that among footprints for which a model assigns, for example, a 70\% blend probability, those footprints should indeed be blended about 70\% of the time. Figure~\ref{fig:calibration_comparison} illustrates the reliability of this interpretation, for Pipeline footprints. Calibration points below the black line indicate a systematic overestimate of the blend probability relative to the fraction of test footprints that are actually blended, and points above the line indicate a corresponding underestimate. Every model is relatively likely to assign classification scores with relatively high certainty (probabilities greater than 80\% or less than 20\%), reflected in the small (non-visible) vertical error bars in the model probability bins near 0 and 1. For these high-certainty assignments, the Gaussian process model is consistently accurate, as shown by the low vertical placement of the blue line on the far left and the high placement on the far right. For probabilities below 20\% and near 65\%, this model is relatively well-calibrated, as shown by the proximity of the blue line to the black line. Elsewhere the model is still generally accurate but underconfident, with blend probabilities higher than they should be but still less than 50\% for many unblended examples (the undershoot of the blue line between 10\% and 50\%), and probabilities lower than they should be but still greater than 50\% for many blended examples (the overshoot of the blue line above 70\%).

In contrast, while the logistic regression model is relatively well-calibrated for probability scores between 50\% and 90\%, its other scores are consistently overestimates of the blend probability. In particular, it is never willing to assign a blend probability less than 10\%, leaving the leftmost bin empty in the figure. The CNN model appears everywhere to exhibit an opposing tendency to the Gaussian process model, assigning overconfident scores across the scale. On the one hand, this is an effect of the highly smooth kernel structure used for the Gaussian process model, resulting in relatively nearby points in PCA embedding space often being assigned very similar class predictions (as remarked above), requiring a conservative estimate of blend probability in order to minimize the validation set cross-entropy. On the other hand, this is also an effect of the high fitting capacity of the CNN model, which can essentially ``memorize'' its limited training set and tweak its probability estimates so that they are extremely close to 0 or 1 on training examples in order to minimize training set cross-entropy. Even though the CNN model has a very low test set cross-entropy, which is consistent with the fact that it was trained to minimize cross-entropy, the calibration plot shows that this can be achieved without necessarily having the most well-calibrated probabilities. Among those examples for which it assigns around a 60\% blend probability, the CNN model is relatively well-calibrated, but the large vertical error bars attest to the fact that this specific probability range is not often assigned. Rather, it almost exclusively assigns probability estimates greater than 90\% or less than 10\%. These are generally accurate as class estimates, but as the plot shows, they are not accurate quite as often as the extreme scores suggest.

\begin{figure}
\gridline{\fig{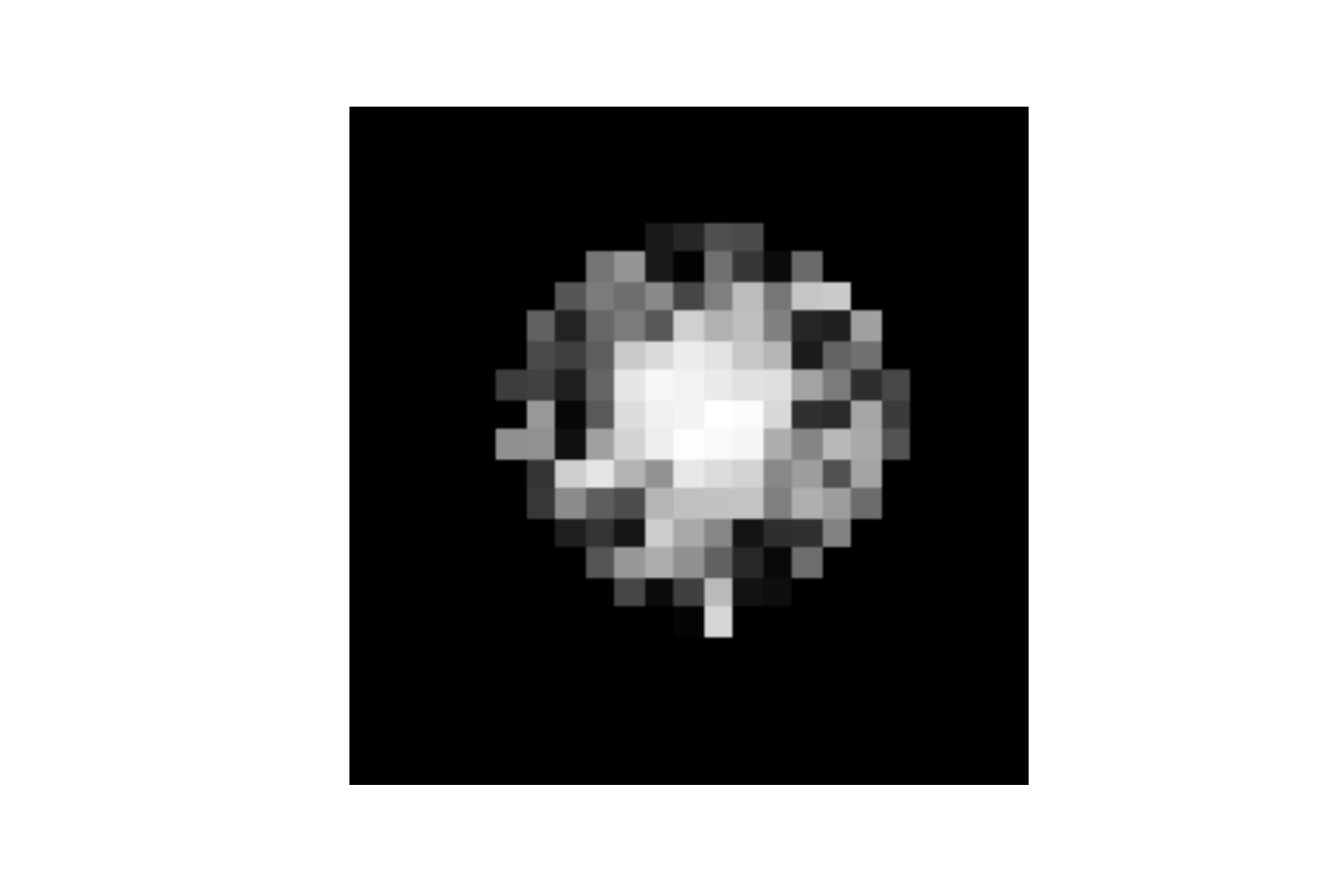}{0.3\textwidth}{(a) Unblended}
          \fig{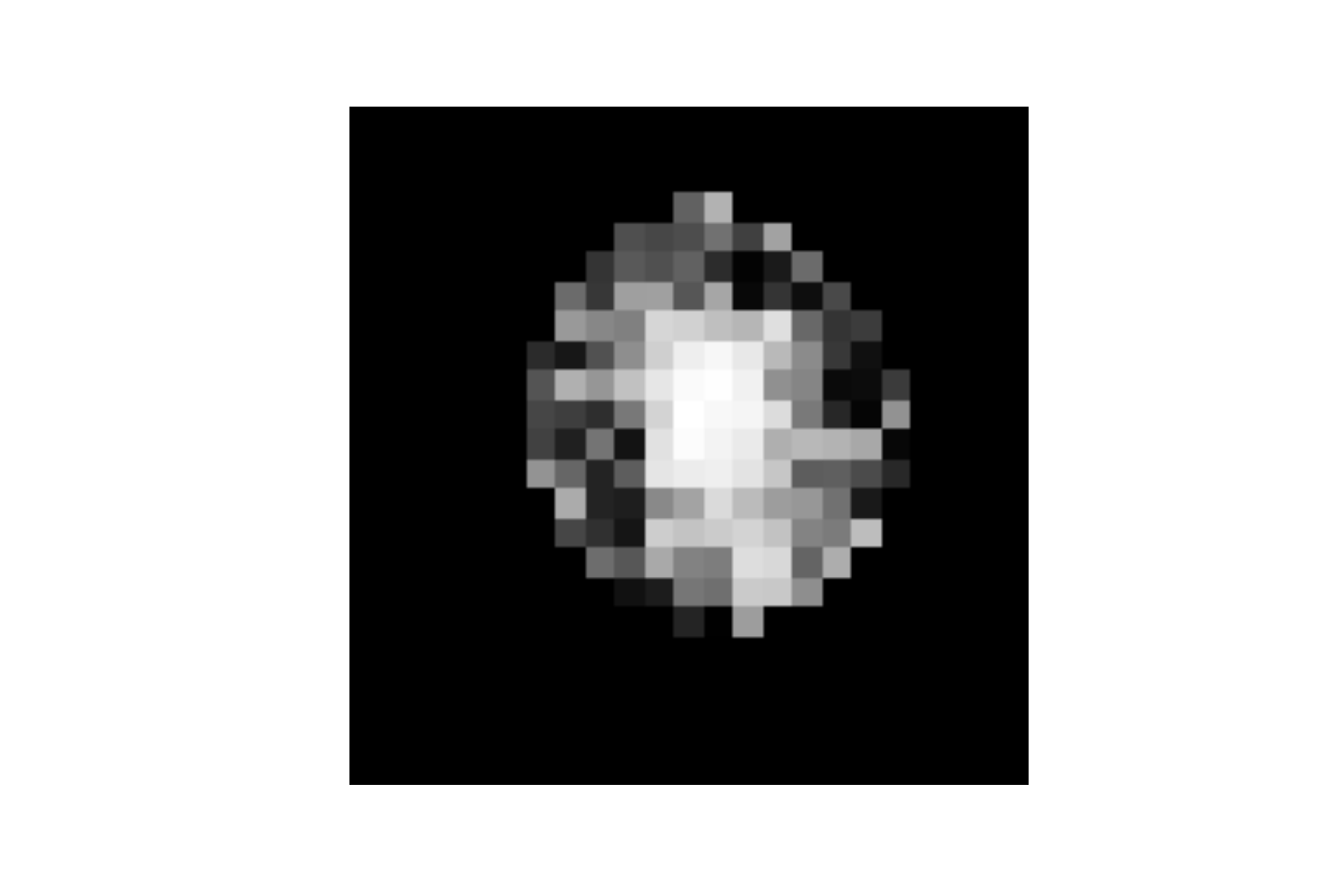}{0.3\textwidth}{(b) Blended}
          \fig{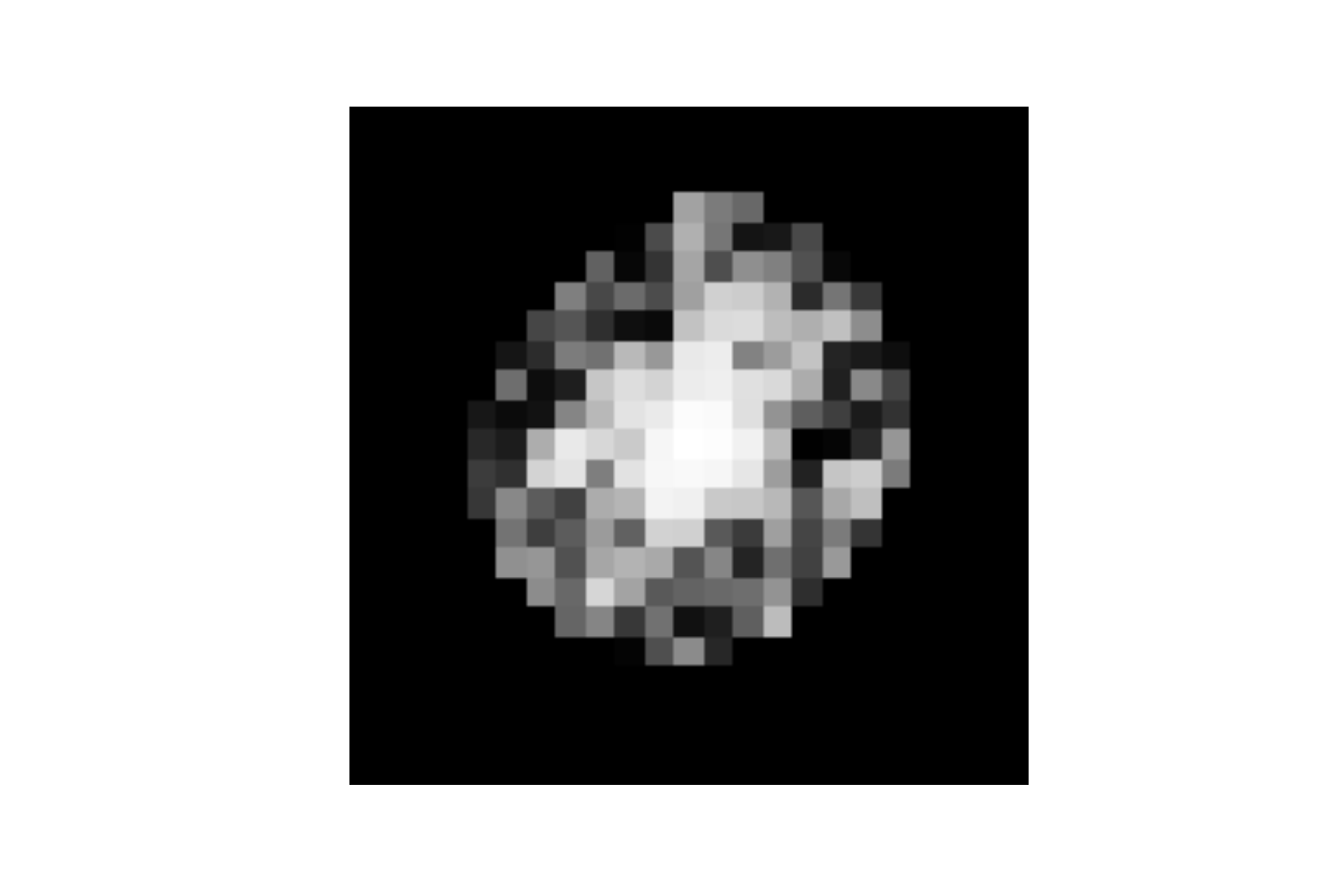}{0.3\textwidth}{(c) Unblended}}
\gridline{\fig{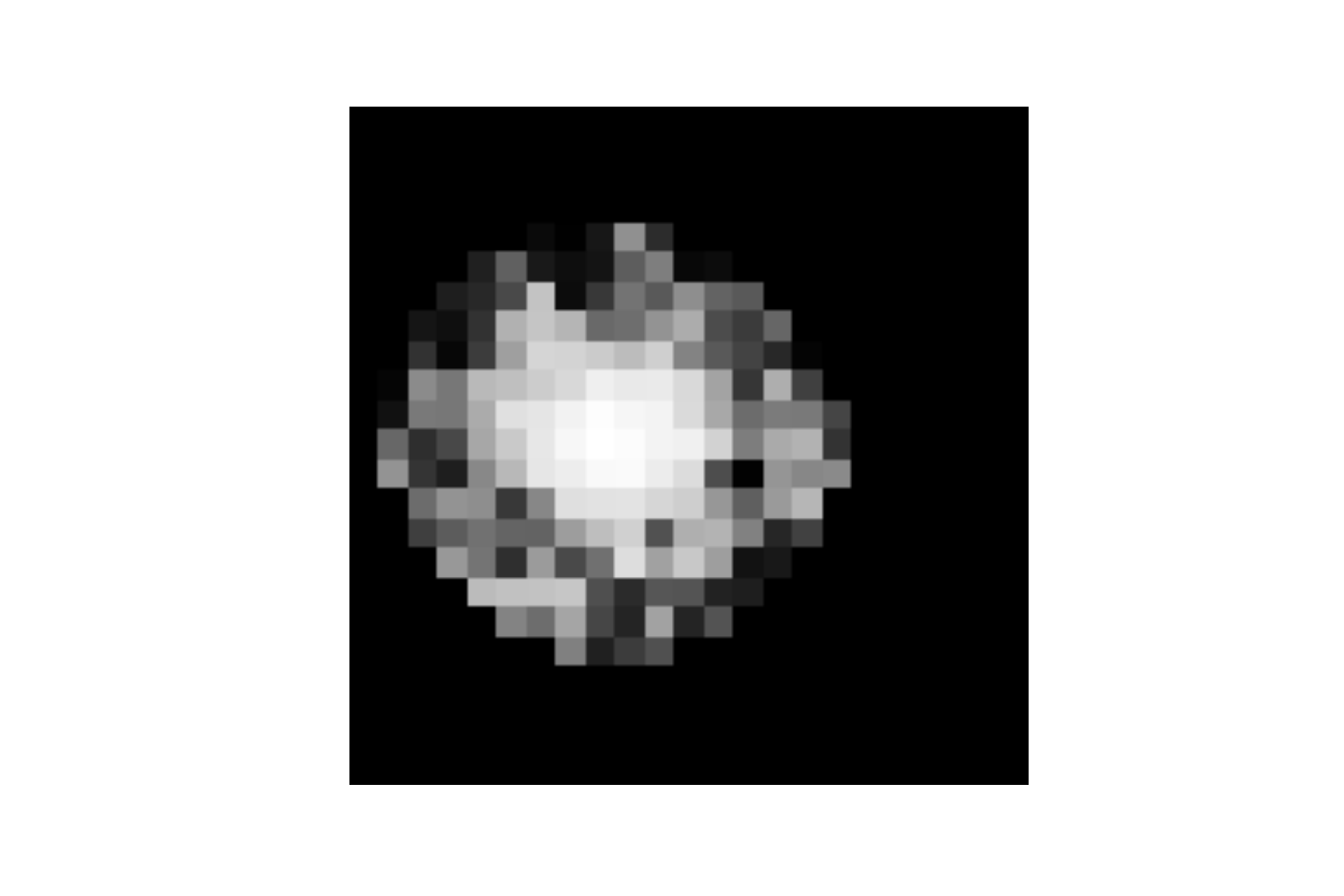}{0.3\textwidth}{(d) Blended}
          \fig{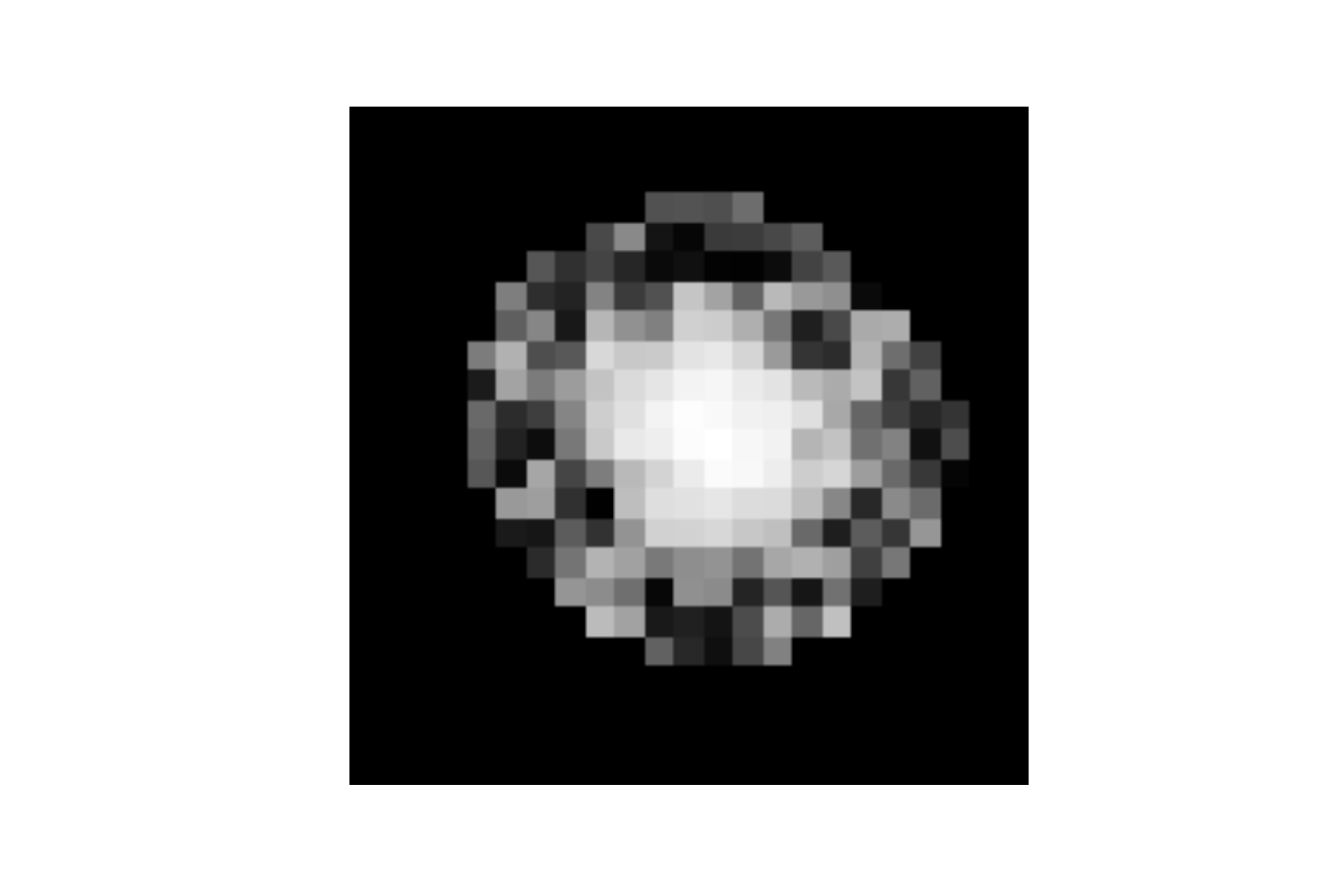}{0.3\textwidth}{(e) Unblended}
          \fig{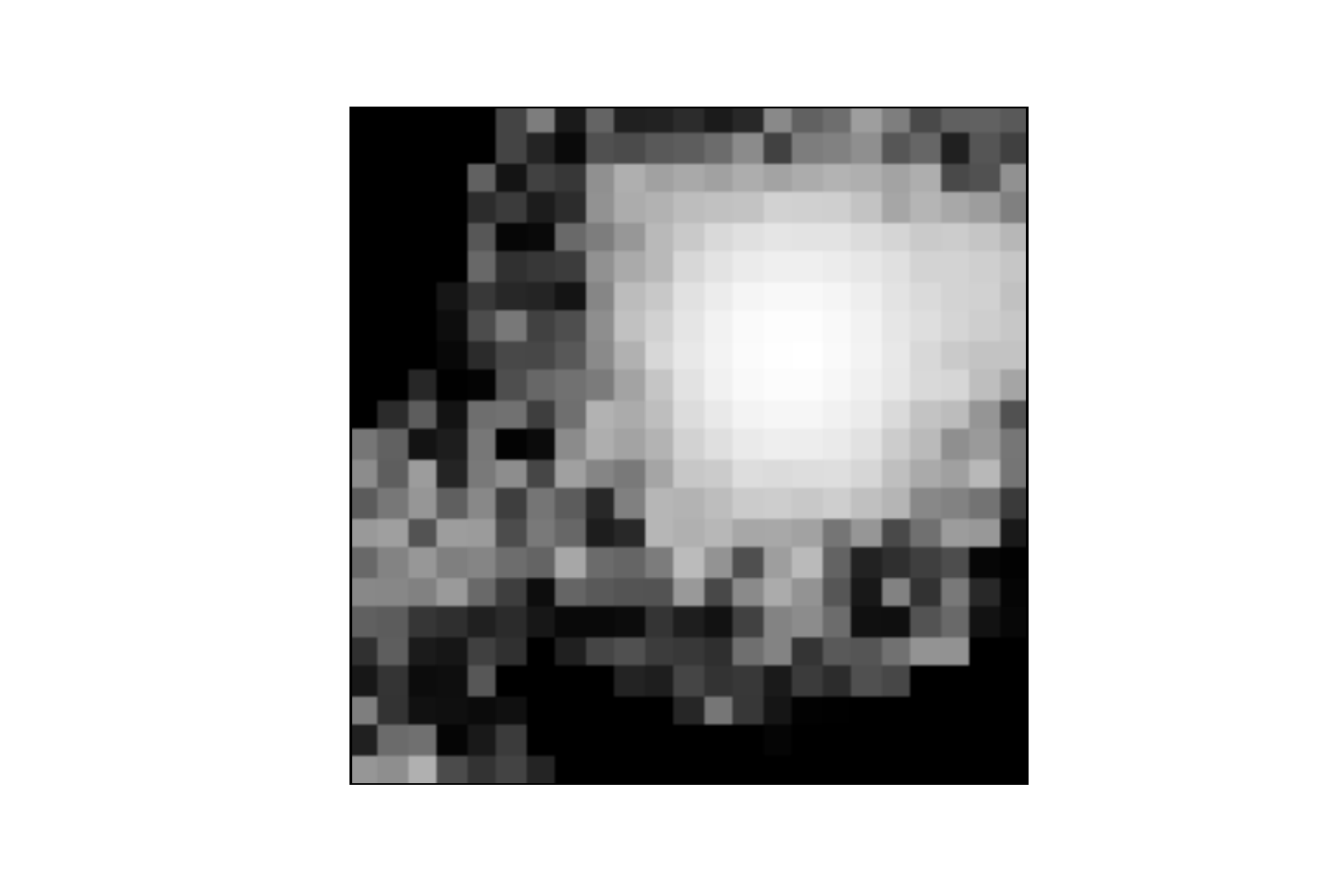}{0.3\textwidth}{(f) Blended}}
\caption{A selection of cutouts.\label{fig:cutouts}}
\end{figure}

Figure~\ref{fig:cutouts} shows a selection of cutouts arranged in order of increasing brightness. Each cutout is a 23x23 pixel image showing the coadd pixels corresponding to one footprint in the test set, with the center pixel overlapping that footprint's intensity barycenter; all pixels not part of the footprint have been blacked out. Thus, these images correspond to the raw inputs to the probabilistic models (prior to PCA embedding, in the case of the logisitic regression and Gaussian process models). Table~\ref{tab:cutout_scores} shows the probability scores assigned to each of these cutouts by specific trained instances of each of the probabilistic models, along with the number of peaks in the smoothed image of each footprint, and the true label (blended or unblended). This table highlights certain features of the Gaussian process and CNN models noted above: the Gaussian process model generally makes more conservative predictions, while the CNN model generally makes highly certain predictions, whether or not they are correct or in agreement with the other models.
\begin{deluxetable*}{cccDDD}
\tablecaption{The truth label, number of peaks, and model-assigned probability scores for each of the footprints in Figure~\ref{fig:cutouts}.\label{tab:cutout_scores}}
\tablehead{\colhead{Cutout} & \colhead{Truth} & \colhead{Peaks} & \twocolhead{GP} & \twocolhead{CNN} & \twocolhead{LR}}
\decimals
\startdata
(a) & Unblended & 1 & 0.052 & 0.953 & 0.246 \\
(b) & Blended   & 1 & 0.471 & 0.401 & 0.255 \\
(c) & Unblended & 1 & 0.510 & 0.147 & 0.341 \\
(d) & Blended   & 1 & 0.307 & 0.939 & 0.346 \\
(e) & Unblended & 2 & 0.283 & 0.964 & 0.401 \\
(f) & Blended   & 4 & 0.957 & 7.46\text{e}{-11} & 0.974
\enddata
\end{deluxetable*}

For example, the CNN model is the only method that correctly classifies footprint (d), assigning a blend probability of 93.9\% to this blended footprint. The fact that this footprint appears off center is an artifact of its very close proximity to the edge of the full simulated scene\footnote{In edge cases such as these, we have kept the cutout size fixed at 23x23 but not allowed the cutout borders to extend past the full scene borders. Otherwise, in this study we have included these footprints in the training and testing sets with no further special handling. Objects this close to the edge of a patch (the nearest equivalent in the LSST Science Pipelines to our simulated ``scenes'') will also be contained in a neighboring patch a hundred or more pixels further from its edge. Hence, in practice most of these edge cases can be safely ignored, though this is a topic of ongoing study.}. Peak finding is clearly insensitive to the location of a footprint in the image (so long as it's not significantly cut off at the edges), and CNNs are approximately translation-invariant by design, but the PCA embedding we have performed here is not, which may contribute to the failure of the logistic regression and Gaussian process models in this case. Conversely, though footprint (f) is clearly blended, has a very wide profile, multiple peaks, and is assigned very high blend scores by all the other models, the CNN model still gives it less than a one in ten billion chance of being blended. Such extreme scores, both right and wrong, are not uncommon with this model. In contrast, the Gaussian process model assigns a 51.0\% blend probability to footprint (c), resulting in an incorrect ``best guess'' blend label but with a very high degree of uncertainty. The Gaussian process model also correctly classifies footprint (e) with fairly high confidence, even though it has two peaks and the CNN model gives it a 96.4\% chance of being blended.

Aside from $\sigma$, the remaining hyperparameters of the Gaussian process model were optimized with respect to classification accuracy, as opposed to the cross-entropy loss. It is possible that the Gaussian process model's probability estimates are not as reliable as they could be if, instead, all hyperparameters were simultaneously optimized with respect to the cross-entropy loss. However, since cross-entropy optimization does not directly target classification accuracy, a corresponding model might classify examples incorrectly more often. Here we have attempted to produce a model that is simply as accurate as possible. The potential tradeoffs between classification accuracy and probabilistic calibration remain a topic for future study, and the decision to prioritize one over the other ultimately depends on a specific use case.

\section{Conclusions and Outlook}\label{sec:conclusions}
We have studied the reliability of galaxy blend identification strategies on \textit{i}-band images from the LSST. This was enabled by a comprehensive simulation of how tens of thousands of galaxies would appear in LSST images. Since these simulated galaxies have a realistic distribution of properties, drawn from the cosmoDC2 catalog, the performance estimates obtained here should be a fair approximation of blend identification performance on real telescope images. Peak finding, the current LSST Science Pipelines strategy, assigns the correct blend ID to about three quarters of all footprints constructed by the current Pipelines. This performance depends sharply on the precise footprint construction method. On average, the Gaussian process model developed here outperforms peak finding on Pipeline-style footprints in almost all contexts, and is substantially less sensitive to the footprint construction method. The image simulation, preprocessing, and model evaluation code used in this study is open source under the MIT License and available on GitHub at \href{https://github.com/jjbuchanan/gp_blendclass_singleband}{jjbuchanan/gp\_blendclass\_singleband}.

A footprint with only one peak that is still classified as blended by a second algorithm is likely to be especially difficult to deblend accurately, and so at a minimum the Gaussian process model could be used to flag these ``bad'' blends. However, given its consistently reliable performance, it appears reasonable to simply prioritize the Gaussian process's predictions over peak finding in all cases. Because the CNN classifier had a higher unblended accuracy than the Gaussian process classifier, while the CNN's blended accuracy was generally lower than the Gaussian process's, it may be possible to combine the two methods in a way that cancels out some classification mistakes and outperforms either single model on its own. However, a combination of models obscures the probabilistic interpretation that we have studied here for each individual model.

The present analysis of classifier performance holds true in \textit{i}-band images where galaxies are the only significant sources of light and variability (together with simplified models of the sky background and PSF). In future work we can take advantage of the images produced by the full DESC DC2 image simulation effort in order to probe the impact of other significant sources of light, such as Milky Way stars, and more realistic modeling of optical variance. The ultimate test will be on real telescope data, such as produced by the HSC, although for real images it will be more difficult to accurately identify which footprints are and are not truly blended. For this purpose, it may be possible to take advantage of the deeper, but narrower, view provided by a space telescope pointed at the same sky coordinates, as in \citet{Dawson_2015} and \citet{2021_MuyGPs_StarGalaxy}. Owing to the relative insensitivity of Gaussian process classification to the specifics of footprint construction, we speculate that the Gaussian process model will maintain very similar performance as more image realism is added. Similarly, we anticipate that a Gaussian process model trained on realistic simulated images will perform comparably well on real telescope images.

The HSC and LSST Pipelines use peak finding not only in single-filter bands, but also in composite footprints formed across all available bands. We have yet to study the reliability of peak finding on these multi-band footprints, or indeed any other single band besides the \textit{i}-band. It remains to be seen whether, and how, incorporating other filter bands in the Gaussian process model could improve its performance still further. This might be done by concatenating bandwise embeddings, as in \citet{2021_MuyGPs_StarGalaxy}. The specific Gaussian process implementation we use assumes a prior mean of 0 for every possible input, but in fact we have seen that brighter footprints are several times more likely to be blended, a priori, than dimmer footprints. We speculate that a more sophisticated treatment of the prior mean could improve the Gaussian process model's performance at different magnitudes, and in particular might shore up its relatively low unblended accuracy at lower magnitudes. This may simply require balancing the class representation in each of several different magnitudes during training, as opposed to just balancing the overall numbers of each class as in the present study.

The Gaussian process model naturally assigns probabilistic estimates to its predictions in a way that peak finding does not. However, the Gaussian process model on its own does not provide a direct estimate of galaxy positions, which is especially significant if deblending is to be attempted. For peak counting to be replaced, some other method would have to be developed for galaxy localization. One strategy would be to train an additional model to predict the positional coordinates of all the galaxies thought to be participating in a given blend. The present blend classifier model is limited to classifying one vs.\ more than one galaxy in a footprint, so for this purpose it may be preferable to develop a multi-class classifier that can decide whether e.g.\ two, three, or four or more galaxies are participating in a given blend. This could be done in hierarchical fashion, starting with a one galaxy vs.\ more than one galaxy classifier, followed by two vs.\ more than two for footprints with more than one, followed by three vs.\ more than three for footprints with more than two, etc. Each of these conditional probabilities can be estimated using a Gaussian process model of exactly the kind described here. For each specific number of galaxies, a distinct Gaussian process regressor could be trained to predict galaxy coordinates, which would ultimately give full probability density maps for the positions of every galaxy.

\begin{acknowledgments}
This work was performed under the auspices of the U.S.
Department of Energy (DOE) by Lawrence Livermore National Laboratory (LLNL) under Contract DE-AC52-07NA27344, with IM release number LLNL-JRNL-824617.
Funding for this work was provided as part of the DOE Office of Science, High Energy Physics cosmic frontier program, and by LLNL Laboratory Directed Research and Development grant 19-SI-004.

This document was prepared as an account of work sponsored by an agency of
the United States government. Neither the United States government nor Lawrence
Livermore National Security, LLC, nor any of their employees makes any warranty,
expressed or implied, or assumes any legal liability or responsibility for the accuracy,
completeness, or usefulness of any information, apparatus, product, or process disclosed,
or represents that its use would not infringe privately owned rights. Reference
herein to any specific commercial product, process, or service by trade name, trademark,
manufacturer, or otherwise does not necessarily constitute or imply its endorsement,
recommendation, or favoring by the United States government or Lawrence Livermore
National Security, LLC. The views and opinions of authors expressed herein
do not necessarily state or reflect those of the United States government or Lawrence
Livermore National Security, LLC, and shall not be used for advertising or product
endorsement purposes.
\end{acknowledgments}

\software{MuyGPs \citep{2021_MuyGPs},
        GCRCatalogs \citep{2018APJS.234.2},
        GalSim \citep{2015AandC.10},
        Source Extractor \citep{1996AAPS.117},
        SEP \citep{2016JOSS.1.6},
        LSST Science Pipelines (\href{https://pipelines.lsst.io}{pipelines.lsst.io})
        }

\bibliography{draft}{}
\bibliographystyle{aasjournal}

\end{document}